# Selected Topics in Asynchronous Automata

Serban E. Vlad
Oradea City Hall & The University of Oradea
str. Zimbrului, Nr.3, Bl.PB68, Et.2, Ap.11, 3700
Oradea, România, serbanvlad@excite.com

***Abstract*** *The paper is concerned with the modeling of the electrical signals and of the asynchronous circuits by* $\mathbf{R} \to \{0,1\}$ *functions and asynchronous automata. The equations of the asynchronous automata are written and solved; the stability, the fundamental mode of operation, the semi-modularity and some aspects of semantics of the propositional temporal logic as seen from behind asynchronous automata are presented.*

**Contents**

## 1. Introduction

The asynchronous circuits, also called Boolean circuits or digital circuits, are formed of logical gates and wires. Their models are called asynchronous automata. The asynchronous circuits may be identified, from a logical point of view, with two families of input, respectively output electrical signals + the relations of determinism between them. Similarly, the asynchronous automata may be identified with two families of $\boldsymbol{R} \to \{0,1\}$ functions, called input and state (or output) functions, that model the input and the output electrical signals + the relations of determinism between them. We have called these relations the equations of the asynchronous automata.

Our main purpose was that of giving a model for the asynchronous circuits and this research was made along many years of informational isolation, the first results being presented in [Vlad, 1989] and [Vlad, 1992]. The terminology from there is that of the fields theory, as resulted by following analogies, in the first case, respectively that of the automata theory, as this theory was presented in the books of Grigore Moisil from the 50's and the 60's referring to the schemata with contacts and relays, in the second case. It has become obvious that, independently on the wish of modeling, the study of the $\boldsymbol{R} \to \{0,1\}$ functions is interesting by itself and it has lead to derivatives, integrals, measures and distributions and we have published some papers on such topics in the "Analele Universitatii din Oradea", some time ago.

The new era of the Internet represents an informational blow-up: let us just mention [Verhoeff, Peeters, 1999], where we have found about 1200 titles of papers concerning asynchronous automata, many of these papers being available. When reading, we have rediscovered the intuition that has come to us since the years of studentship from our professors, that has been the main source of inspiration - Timisoara, electrical engineering - and we have put order in the already existing results. The conclusion: there does not exist a mathematical theory of the asynchronous automata.

These were the circumstances in which the present paper was written, with the desire of giving intuition - we have often quoted due to this reason the bibliography - and of formalizing afterwards, as much as possible. This type of connection to the bibliography should suggest the distance, sometimes long, sometimes short, from between electrical engineering and mathematics.

The main ideas that we deal with in the paper refer to modeling, writing and solving the equations of the asynchronous automata, defining and characterizing the synchronous-like automata, i.e. the automata with semi-modular transitions and the fundamental mode of operation and to presenting some aspects of semantics of the temporal logic, as seen from behind systems theory.

We express our gratitude for the support and friendship shown by Prof. Dr. Luciano Lavagno from Udine, that has had long debates on asynchronous topics with us in the period when the paper was written.

## 2. Preliminaries

2.1 **Definition** The *Boole algebra with two elements*, or the *Boolean ring with two elements* $\boldsymbol{B}_2$ consists in the set $\{0,1\}$ together with

a) the order $0 \le 1$

b) the discrete topology (i.e. the open sets are the subsets of $\{0,1\}$)

c) the laws: the *logical complement* '$-$', the *modulo 2 sum* '$\oplus$' and the *product* '$\cdot$'

| $-$ | 0 | 1 |
|---|---|---|
| | 1 | 0 |

a)

| $\oplus$ | 0 | 1 |
|---|---|---|
| 0 | 0 | 1 |
| 1 | 1 | 0 |

b)

| $\cdot$ | 0 | 1 |
|---|---|---|
| 0 | 0 | 0 |
| 1 | 0 | 1 |

c)

table (1)

2.2 **Remark** It is obvious expressing the *reunion* '$\cup$' with $\oplus$, $\cdot$ and proving that $\boldsymbol{B}_2$ is a Boole algebra relative to $\bar{\ }$, $\cup$, $\cdot$. $\boldsymbol{B}_2$ is also a field, in particular a ring, relative to $\oplus$, $\cdot$.

2.3 **Remark** In the rest of this paragraph we shall consider that there are given a function $x:\boldsymbol{R}\to\boldsymbol{B}_2$ and the subset $I\subset\boldsymbol{R}$; special cases: $I=\boldsymbol{R}$ and $I=\emptyset$.

2.4 **Definition** $\chi_I:\boldsymbol{R}\to\boldsymbol{B}_2$ is the *characteristic function* of the set $I$:

$$\chi_I(t)=\begin{cases}1, t\in I\\ 0, t\notin I\end{cases} \quad (1)$$

2.5 **Definition** The *support* of $x$ is the set:

$$supp\ x=\{t\,|\,t\in\boldsymbol{R}, x(t)=1\} \quad (1)$$

2.6 **Remark** 2.4 and 2.5 give

$$supp\ \chi_I=I \quad (1)$$

$$x=\chi_{supp\ x} \quad (2)$$

2.7 **Notation** We shall note with $|I|$ the number of elements of the set $I$, supposing that it is finite. If $I$ is infinite, $|I|$ is not defined.

2.8 **Definition** The *modulo 2 summation* '$\Xi$', the *reunion* '$\bigcup$' and the *intersection* '$\bigcap$' *of* $x$ *on* $I$ are given by:

$$\underset{t\in I}{\Xi}\ x(t)=\begin{cases}1, |I\wedge supp\ x|\ is\ odd\\ 0, |I\wedge supp\ x|\ is\ even\end{cases}, \qquad \underset{t\in\emptyset}{\Xi}\ x(t)=0 \quad (1)$$

$$\bigcup_{t\in I}x(t)=\begin{cases}1, I\wedge supp\ x\neq\emptyset\\ 0, I\wedge supp\ x=\emptyset\end{cases}, \qquad \bigcup_{t\in\emptyset}x(t)=0 \quad (2)$$

$$\bigcap_{t\in I}x(t)=\begin{cases}1, supp\ x=I\\ 0, supp\ x\neq I\end{cases}, \qquad \bigcap_{t\in\emptyset}x(t)=1 \quad (3)$$

where '$\wedge$' is the meet (or the intersection) of the sets and at (1), $I\wedge supp\ x$ is supposed to be finite.

2.9 **Remark** $\underset{t\in I}{\Xi}\ x(t)$ is a generalized series; the series is convergent if $I\wedge supp\ x$ is finite and divergent otherwise. $\bigcup_{t\in I}x(t)$ is the maximum of the function *x* on the set *I* and $\bigcap_{t\in I}x(t)$ is the minimum of the function *x* on the set *I*.

We have considered that $|\emptyset|=0$ is an even number.

2.10 **Definition** The *left limit* and the *right limit* $x(t-0), x(t+0),$ respectively the *left derivative* and the *right derivative* $D^{-}x(t), D^{+}x(t)$ of $x$ in $t$ are binary numbers or binary functions, as $t$ is fixed or variable and they are defined like this:

$$\exists\varepsilon > 0, \forall\xi\in (t-\varepsilon, t), x(\xi) = x(t-0) \quad (1)$$

$$\exists\varepsilon > 0, \forall\xi\in (t, t+\varepsilon), x(\xi) = x(t+0) \quad (2)$$

$$D^{-}x(t) = x(t-0)\oplus x(t) \quad (3)$$

$$D^{+}x(t) = x(t+0)\oplus x(t) \quad (4)$$

2.11 **Definition** A function $x$ for which $x(t-0)$ and $x(t+0)$ exist for all $t\in \boldsymbol{R}$ is called *differentiable*.

2.12 **Notations** We note with $Diff$ the set of the differentiable functions and with $Diff^{(n)}$ the set

$$Diff^{(n)} = \{(x_1(t),...,x_n(t)) \mid t\in \boldsymbol{R}, x_i \in Diff, i = \overline{1,n}\} \quad (1)$$

2.13 **Remark** The relation between $Diff^{n}$ and $Diff^{(n)}$ is the following one. Let $i: \boldsymbol{R}\to \boldsymbol{R}^{n}$ defined by

$$\boldsymbol{R}\ni t\mapsto i(t) = (t,...,t)\in \boldsymbol{R}^{n} \quad (1)$$

Then

$$Diff^{(n)} = \{x\circ i \mid x\in Diff^{n}\} \quad (2)$$

2.14 **Remarks** The previous notions (derivatives, differentiability) have formal similarities with those of the real functions. These similarities justify the terminology.

On the other hand if $D^{-}x(t) = 0$, then $x$ is left continuous in $t$ and if $D^{-}x(t) = 1$, then $x$ switches (from 0 to 1 or from 1 to 0): it is left discontinuous in $t$.

2.15 **Proposition** $Diff$ is a ring relative to the laws $\oplus, \cdot$ that are induced by those of $\boldsymbol{B}_2$ and $Diff^{(n)}$ is a $\boldsymbol{B}_2$-linear space relative to the obvious laws $\oplus, \cdot$.

2.16 **Theorem** (of representation of the differentiable functions). The following statements are equivalent:

a) $x\in Diff$

b) the real numbers $t_z \in \boldsymbol{R}$ and the binary numbers $a_z, b_z \in \boldsymbol{B}_2, z\in \boldsymbol{Z}$ exist so that:

$$... < t_{-1} < t_0 < t_1 < ... \quad (1)$$

$$\forall t', t''\in \boldsymbol{R}, (t', t'')\wedge \{t_z \mid z\in \boldsymbol{Z}\} \text{ is finite} \quad (2)$$

$$x(t) = ...\oplus a_{-1}\cdot\chi_{\{t_{-1}\}}(t)\oplus b_0\cdot\chi_{(t_{-1},t_0)}(t)\oplus a_0\cdot\chi_{\{t_0\}}(t)\oplus b_1\cdot\chi_{(t_0,t_1)}(t)\oplus ... \quad (3)$$

2.17 **Remarks** From the proof - that is omitted - of the previous theorem, we just mention that for any $t\in \boldsymbol{R}$:

$$x(t-0) = b_z, \text{ where } z \text{ is chosen so that } t\in (t_{z-1}, t_z] \quad (1)$$

$$x(t+0) = b_z, \text{ where } z \text{ is chosen so that } t\in [t_{z-1}, t_z) \quad (2)$$

We also observe that, given $x$, the families $(t_z), (a_z), (b_z)$ are not unique (in order to see this, we can take $x$ to be the constant function).

2.18 **Remark** The theorem 2.16 shows that the differentiable functions are compatible with the electrical signals of the inertial digital circuits. The abstract model of such a circuit is called asynchronous automaton and it makes use of differential equations written with differentiable functions.

2.19 **Definition** Let the real family $(t_z)$. It is called *strictly increasing,* if it fulfills the condition 2.16 (1); *locally finite,* if it fulfills 2.16 (2), respectively *strictly increasing locally finite,* shortly SILF, if it fulfills 2.16 (1) and 2.16 (2).

2.20 **Remark** For $x$ like at 2.16 (3), simple computations show that:

$$x(t-0) = ... \oplus b_0 \cdot \chi_{(t_{-1},t_0]}(t) \oplus b_1 \cdot \chi_{(t_0,t_1]}(t) \oplus ... \quad (1)$$

$$D^- x(t) = ... \oplus (a_0 \oplus b_0) \cdot \chi_{\{t_0\}}(t) \oplus (a_1 \oplus b_1) \cdot \chi_{\{t_1\}}(t) \oplus ... = \underset{z \in \mathbf{Z}}{\Xi} D^- x(t_z) \cdot \chi_{\{t_z\}}(t) \quad (2)$$

thus $supp\ D^- x \subset (t_z)$. The situation is similar for the right statements.

2.21 **Remark** An interesting consequence of these ideas is: at the left of an arbitrary point $t$, the derivative of $x$ is null:

$$D^- x(t-0) = 0 = x((t-0)-0) \oplus x(t-0) \quad (1)$$

from where

$$x((t-0)-0) = x(t-0) \quad (2)$$

2.22 **Remark** There do not exist distinct lateral limits of the second order and distinct derivatives of the second order. There does not exist a differentiability of the second order of the functions $\mathbf{R} \to \mathbf{B}_2$.

2.23 **Definition** The set $Real \subset Diff$ of the *realizable* functions is defined by:

$$supp\ x \subset [0, \infty) \quad (1)$$

$$supp\ D^+ x = \emptyset \text{ (i.e. } D^+ x(t) = 0, t \in \mathbf{R}) \quad (2)$$

and the set $Real^{(n)} \subset Diff^{(n)}$ is given by:

$$Real^{(n)} = \{(x_1(t),...,x_n(t)) \mid t \in \mathbf{R}, x_i \in Real, i = \overline{1,n}\} \quad (3)$$

2.24 **Proposition** $Real$ is a subring of $Diff$ and $Real^{(n)}$ is a linear subspace of $Diff^{(n)}$.

2.25 **Definition** A real family $(t_z)$ satisfying

$$0 = t_0 < t_1 < ... \quad (1)$$

and 2.16 (2) is called *strictly increasing non-negative locally finite,* shortly SINLF.

2.26 **Proposition** (of representation of the realizable functions). The following statements are equivalent:

a) $x \in Real$

b) the SINLF family $t_z \in \mathbf{R}, z \in \mathbf{Z}$ exists so that:

$$x(t) = x(t_0) \cdot \chi_{[t_0,t_1)}(t) \oplus x(t_1) \cdot \chi_{[t_1,t_2)}(t) \oplus ... \quad (1)$$

2.27 **Remarks** The differentiable functions are self-dual.

The realizable functions and respectively their right duals, the realizable* functions, model the behavior of the deterministic (i.e. non-anticipative) circuits, where the present depends on the past but not on the future, respectively of the anticipative circuits, where the present depends on the future but not on the past. The intuitive meaning of left and right is given, from a systemic point of view, by the duality determinism - anticipativity.

The realizable functions point out the existence of the initial time moment 0, whilst the realizable* functions point out the existence of the final time 0.

2.28 **Definition** The set $Diff_+ \subset Diff$ of the *differentiable functions with a non-negative support* is defined by:

$$supp\ x \subset [0,\infty) \tag{1}$$

and the set $Diff_+^{(n)} \subset Diff^{(n)}$ is given by:

$$Diff_+^{(n)} = \{(x_1(t),...,x_n(t)) \mid t \in \boldsymbol{R}, x_i \in Diff_+, i = \overline{1,n}\} \tag{2}$$

2.29 **Proposition** We have the next inclusions $Real \subset Diff_+$, $Real^{(n)} \subset Diff_+^{(n)}$ of rings, respectively of linear spaces.

## 3. The Modeling of the Electrical Signals

3.1 **Terminology** Let us consider the next drawing

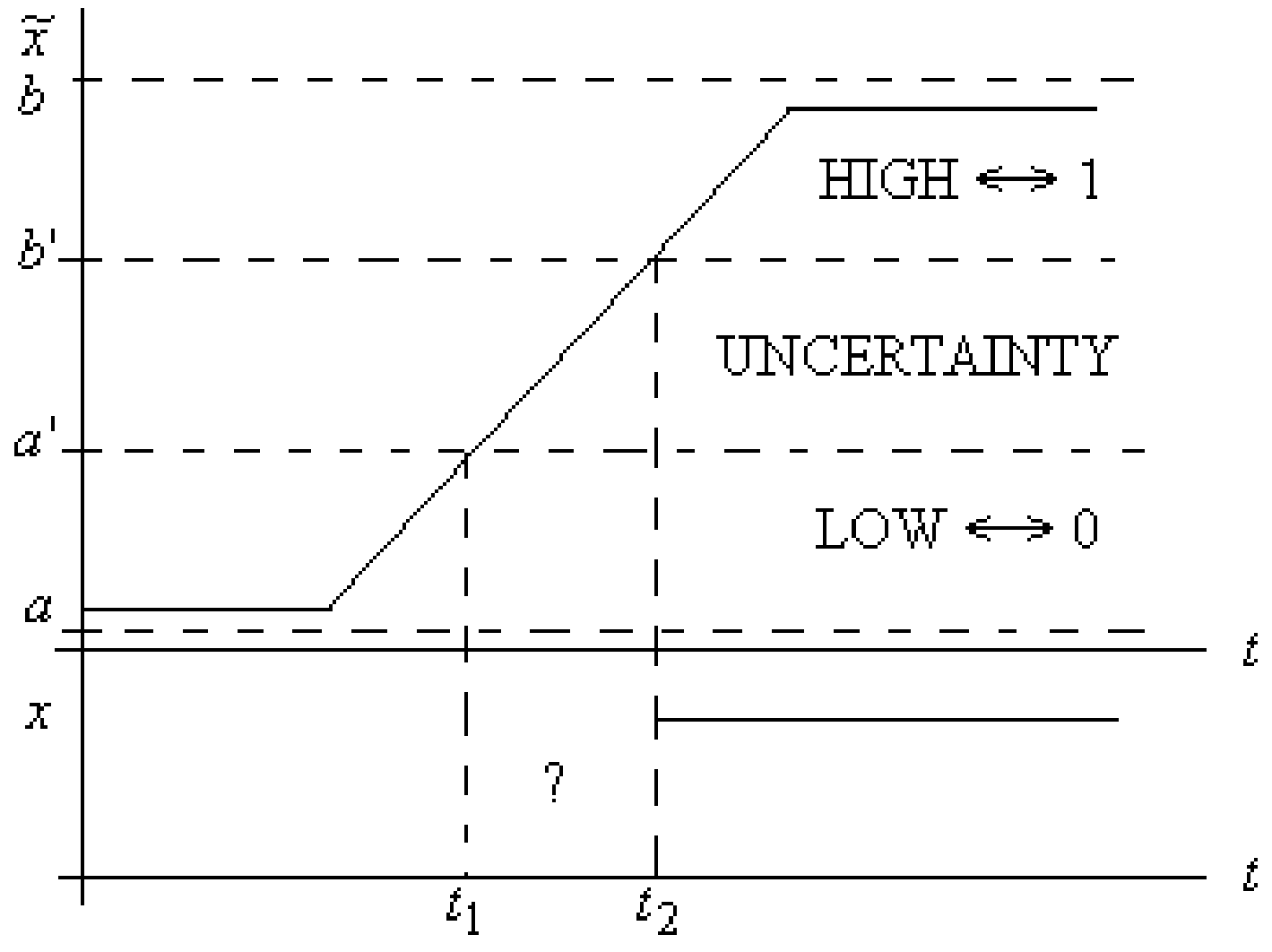

fig (1)

where $0 < a < a' < b' < b$ and

a) $t \in \boldsymbol{R}$ is the *temporal variable*

b) the function $\tilde{x} : \boldsymbol{R} \to [0,b]$ is called *electrical signal*

c) $x : \boldsymbol{R} \to \boldsymbol{B}_2$ is the *binary model* of $\tilde{x}$

d) the intervals $[a,a')$, $[a',b']$, $(b',b]$ are called the ranges of values *LOW, UNCERTAINTY* and *HIGH.*

3.2 **Remark** In figure 3.1 (1) we observe the existence of the time interval $[t_1,t_2]$ when $x$ can take any value, 0 or 1. This situation was noted with the sign ?.

3.3 **Remarks** The electrical signals completely characterize the asynchronous circuits. Such a circuit $\tilde{\Sigma}$ is identified with a finite, non-empty set of electrical signals $\tilde{x}_1,...,\tilde{x}_k$ + the relations of determinism between them.

The $\boldsymbol{R} \to \boldsymbol{B}_2$ functions completely characterize the behavior of the asynchronous automata. Such an automaton $\Sigma$ is identified with a finite, non-empty set $x_1,...,x_k$ of models of the signals $\tilde{x}_1,...,\tilde{x}_k$ + the relations of determinism between them. From this point of view, we can say that $\Sigma$ models $\tilde{\Sigma}$.

We have in fact the next levels of abstractization here:

a) the asynchronous circuit $\tilde{\Sigma}$ - the lowest level of abstractization

b) the electrical signals $\tilde{x}_1,...,\tilde{x}_k$

c) $x_1,...,x_k$ and $\Sigma$ - the highest level of abstractization.

3.4 **Our purpose** in this section is related to 3.3 a), b) c) in the following manner:

a) $\tilde{\Sigma}$ cannot be defined, it is a physical object similar to the tables and the chairs from a living room. What we can do is to call its constitutive elements and to show the way that they can be connected to each other.

b) We shall define the electrical signals.

c) We shall show the meaning of the fact that $x : \boldsymbol{R} \to \boldsymbol{B}_2$ models the electrical signal $\tilde{x}$.

d) We shall show that any electrical signal can be modeled by a realizable function

e) We shall suppose that, when modeling is possible, the realizable functions are models of some electrical signals.

3.5 **Informal definition** "An asynchronous circuit, informally, is an arbitrary interconnection of logic gates, with the only restriction that no two gate outputs can be tied together" (cf. [Lavagno, 1992]).

3.6 **Remarks** By "logic gates" we understand these devices that implement the Boolean functions. The most frequent such gates are related to simple Boolean functions and their symbols have been written bellow:

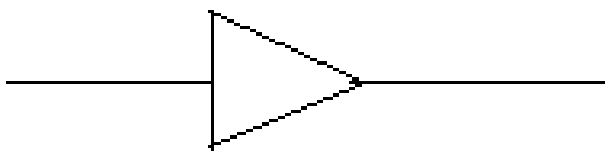

a) delay element; it implements the identical function $f(a) = a$

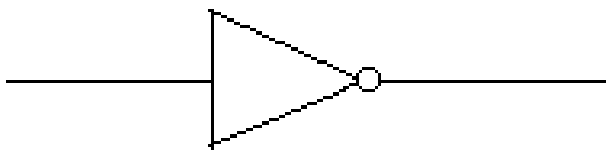

b) the NOT gate; it implements the logical complement function $f(a) = \overline{a}$

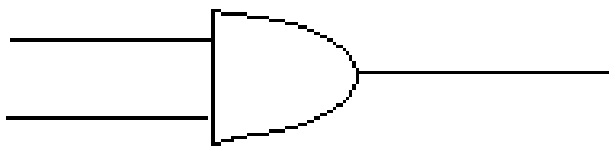

c) the AND gate; it implements the product function $f(a,b) = a \cdot b$

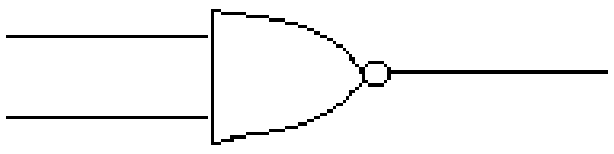

d) the NAND gate; it implements the Scheffer function $f(a,b) = \overline{a \cdot b}$

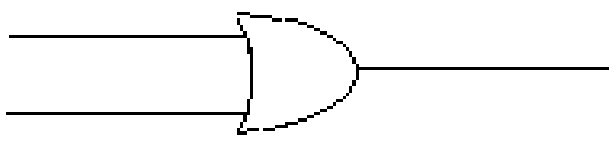

e) the OR gate; it implements the reunion function $f(a,b)=a\cup b$

3.7 **Definition** An *electrical signal* is a function $\tilde{x}:\boldsymbol{R}\to[0,b]$ that associates to each moment of time $t\in\boldsymbol{R}$ (the time is measured in seconds) a value $\tilde{x}(t)$ tension (that is measured in volts). We ask that:

i) $\tilde{x}(t)=0, t<0$

ii) $\tilde{x}(t)\in[a,b], t\geq 0$

iii) $\tilde{x}(t)$ is continuous in all $t\neq 0$; in 0, $\tilde{x}$ is right continuous

iv) $\forall t',t''\in\boldsymbol{R},(t',t'')\wedge\{t\,|\,\tilde{x}(t)\in\{a',b'\}\}$ is finite.

3.8 **Remarks** We interpret the definition 3.7 in the following manner.

a) The requests i), ii) point out that 0 is the initial time moment. This anticipates that the asynchronous automata are constant (or time invariant) dynamical (i.e. determinist, or non-anticipative) systems, cf. with [Kalman+, 1975], definition 1.2 stating roughly that for these systems, the translations along the time axis of the inputs, of the states and of the outputs are possible when they take place simultaneously, with the same $\tau\in\boldsymbol{R}$. Thus, choosing 0 as initial time moment is possible without loss.

b) the requests iii), iv) are of inertiality. To be compared 3.7 iv) with the request of local finiteness 2.16 (2).

3.9 **Definition** The *logical value function* $\nu:[a,a')\vee(b',b]\to\boldsymbol{B}_2$ is defined by:

$$\nu(\xi)=\begin{cases}0, \xi\in[a,a')\\1, \xi\in(b',b]\end{cases} \quad (1)$$

3.10 **Definition** Let $x:\boldsymbol{R}\to\boldsymbol{B}_2$ and $I\subset[0,\infty)$ some set. $x$ is called *model*, or *model function*, or *modeling function of the electrical signal* $\tilde{x}$ (*on the set* $I$) if

$$\forall t\geq 0, \tilde{x}(t)\in[a,a')\vee(b',b]\Rightarrow\nu(\tilde{x}(t))=x(t) \quad (1)$$

$$(\ \forall t\in I, \tilde{x}(t)\in[a,a')\vee(b',b]\Rightarrow\nu(\tilde{x}(t))=x(t)\ ) \quad (2)$$

We say that $x$ *models* the signal $\tilde{x}$ (on $I$).

3.11 **Definition** Let us note

$$I'=\{t\,|\,\tilde{x}(t)\in[a,a')\vee(b',b]\} \quad (1)$$

and we have the next possibilities:

a) $I'=\emptyset$; we say that $\tilde{x}$ *is the trivial electrical signal* and that $x$ *models* $\tilde{x}$ *in the trivial manner.*

b) $I'\neq\emptyset$; we say that $\tilde{x}$ is a *proper* (or *non-trivial*) *electrical signal* and that $x$ *models* $\tilde{x}$ *in a proper* (or *non-trivial*) *manner.*

In the cases a), b) we can replace $I'$ with $I'\wedge I$ when the modeling is done on this set.

3.12 **Definition** We say that $\tilde{x}$ *has a switch* from *LOW-HIGH* or that it *switches* from *LOW-HIGH* if there exist the numbers $0\leq t'<t''$ with $\tilde{x}(t')\in[a,a'), \tilde{x}(t'')\in(b',b]$.

3.13 **Proposition** If $\tilde{x}$ has a switch from *LOW-HIGH* then there exists an interval $[t_1,t_2]\subset(0,\infty)$ with:

i) $\forall\varepsilon>0,\exists\xi\in(t_1-\varepsilon,t_1),\tilde{x}(\xi)\in[a,a')$

ii) $\tilde{x}(t_1)=a'$

iii) $\tilde{x}([t_1,t_2]) = [a',b']$

iv) $\tilde{x}(t_2) = b'$

v) $\forall \varepsilon > 0, \exists \xi \in (t_2, t_2+\varepsilon), \tilde{x}(\xi) \in (b',b]$

**Proof** We have the next property: because $\tilde{x}$ is continuous on $[0,\infty)$ then, for any $0 \le t' < t''$ with $\tilde{x}(t') < \tilde{x}(t'')$ and for any $\lambda \in (\tilde{x}(t'), \tilde{x}(t''))$, there exists at least a $t \in (t',t'')$ so that $\tilde{x}(t) = \lambda$ cf. for example [***,1977], pg. 221.

3.14 **Definition** a) In the conditions of the Proposition 3.13, the interval $[t_1,t_2]$ is called the *switching interval of* $\tilde{x}$ *from LOW-HIGH*. We say that $\tilde{x}$ *has a switch* or that it *switches from LOW-HIGH during the interval* $[t_1,t_2]$.

b) The real positive number $t_2 - t_1$ is called the *switching time of* $\tilde{x}$ *from LOW-HIGH*.

c) Let us suppose that $x$ changes its value in the interval $[t_1,t_2]$ exactly once from $0$ to $1$. We say that $x$ *switches from* $0$ *to* $1$ (*together with* $\tilde{x}$).

3.15 **Remark** From a physical point of view, the switching of the electrical signals is equivalent to the migration of a cloud of electrons inside a semiconductor. The time interval $[t_1,t_2]$ from fig 3.1 (1), Proposition 3.13 and Definition 3.14 is in fact the duration of this migration.

3.16 **Remark** Similar (dual) statements with the previous ones from 3.12, 3.13, 3.14 are obtained by replacing *LOW* with *HIGH* and vice-versa, respectively by replacing $0$ with $1$ and vice-versa.

3.17 **Remark** We have defined in paragraph 2 the realizable functions and the SINLF families. In order to point out what connection exists between the realizable functions and the electrical signals, we observe that, given the signal $\tilde{x} : \boldsymbol{R} \to [0,b]$, we have the next possibilities:

a) $\tilde{x}$ does not switch (from *LOW-HIGH*, or from *HIGH-LOW*) at all

b) $\tilde{x}$ switches (from *LOW-HIGH*, or from *HIGH-LOW*) a finite number of times. There exist the numbers:

$$0 < t_0 < t_1 < ... < t_{2p} < t_{2p+1}$$

with the property that the switching intervals are: $[t_0,t_1],...,[t_{2p},t_{2p+1}]$

c) $\tilde{x}$ switches (from *LOW-HIGH*, or from *HIGH-LOW*) countably many times, in the sense that there exist the real numbers

$$0 < t_0 < t_1 < ... < t_{2p} < t_{2p+1} < ...$$

so that the switching intervals of $\tilde{x}$ are: $[t_0,t_1],...,[t_{2p},t_{2p+1}],...$

3.18 **Proposition** Any electrical signal $\tilde{x}$ has a realizable model $x \in Real$.

**Proof** We refer to the cases a), b), c) from 3.17 when $\tilde{x}$ is modeled by the realizable function $x$ having the form:

$$x(t) = a \cdot \chi_{[0,\infty)}(t) \tag{1}$$

$$x(t) = a \cdot \chi_{[0,t_1)}(t) \oplus \overline{a} \cdot \chi_{[t_1,t_3)}(t) \oplus ... \oplus b \cdot \chi_{[t_{2p-1},t_{2p+1})}(t) \tag{2}$$

$$x(t) = a \cdot \chi_{[0,t_1)}(t) \oplus \overline{a} \cdot \chi_{[t_1,t_3)}(t) \oplus ... \oplus b \cdot \chi_{[t_{2p-1},t_{2p+1})}(t) \oplus ... \tag{3}$$

where $a,b \in \boldsymbol{B}_2, t \in \boldsymbol{R}$ and in (3) it is easily seen that the condition 3.7 iv) implies that $\{t_n \mid n \in \boldsymbol{N}\}$ is SINLF.

3.19 **Convention** In this paper we shall suppose that the electrical signals are modeled by realizable functions.

## 4. Some Words on the Dual Notions

4.1 **Remarks** Let us recall some of the dual concepts that have occured in the paper:

a) the Boolean intersection and reunion; the set intersection and reunion

b) the left limit $x(t-0)$ and the right limit $x(t+0)$ of $x$

c) the left derivative $D^{-}x(t)$ and the right derivative $D^{+}x(t)$ of $x$

d) 0 and 1, *LOW* and *HIGH*

e) the realizable and the realizable* functions

f) the differentiable functions with a non-negative support and the differentiable functions with a non-positive support

g) the past and the future

h) the dynamical, or the non-anticipative systems, respectively the anticipative systems

i) the initial time moment and the final time moment

and there are also similar dual concepts that have not been presented explicitly, for example the modulo 2 sum and the logical value function have dual concepts themselves.

Between the self-dual notions that have occured, we mention:

a') the logical complement

b') the differentiable functions

c') *UNCERTAINTY*

d') the delay element and the identical Boolean function.

The Remark 2.27 shows the way that the dual concepts are related to each other.

The sources that generate dual concepts are, generally:

- of algebraical nature
- related to the duality of the order relations < and > on $\mathbf{B}_2$ and $\mathbf{R}$.

Until now and from now also, understanding duality gives a logical symmetry to the exposure.

## 5. Delays

5.1 **Notation** We consider a delay element whose input electrical signal $\tilde{u}$ and output electrical signal $\tilde{x}$ are modeled by the realizable functions $u, x$ and the delay is the parameter $\tau > 0$.

5.2 **Terminology** The functions $u, x$ are called the *input function*, respectively the *state* (or the *output*) *function.* $\tau$ is called *delay*.

$\tilde{u}, \tilde{x}$ refer to the *delay element* and $u, x, \tau$ refer to the *delay model*.

5.3 **Informal definition**

a) "A delay element is *pure* if it transmits each event on its input to its output, i.e. it corresponds to a pure translation in time of the input waveform", cf. [Lavagno, 1992].

b) "A *pure* delay can delay the propagation of a waveform, but does not otherwise alter it", cf. [Davis, Nowick, 1997].

5.4 **Definition** The delay model and the delay itself are called *pure* if

$$x(t) = u(t-\tau) \tag{1}$$

5.5 **Informal definition**

a) "The *inertial delay model* means that if an element has a switching delay of $\tau$ time units, pulses generated by the logic evaluator with duration less than $\tau$ are filtered out, while pulses longer than $\tau$ units appear at the output $x$ shifted in time by $\tau$ units", cf. [Lavagno, 1992].

b) "An *inertial delay* can alter the shape of a waveform by attenuating short glitches. More formally, an inertial delay has a threshold period $\tau$. Pulses of duration less than $\tau$ are filtered out", cf. [Davis, Nowick, 1997].

5.6 **Definition** The delay model and the delay itself are *inertial* if

$$D^{-}x(t) = (x(t-0) \oplus u(t-0)) \cdot \overline{\bigcup_{\xi \in (t-\tau,t)} D^{-}u(\xi)} \cdot \chi_{[\tau,\infty)}(t) \oplus x^{0} \cdot \chi_{\{0\}}(t) \qquad (1)$$

In the previous equation, the number $x^{0} \in \boldsymbol{B}_2$ is called the *initial state*.

5.7 **Example** of inertial delay model. We take $x^{0} = 0, \tau = 2$ and

$$u(t) = \chi_{[3,4)}(t) \oplus \chi_{[6,9)}(t) \qquad (1)$$

We have:

$$D^{-}u(t) = \chi_{\{3,4,6,9\}}(t) \qquad (2)$$

$$\bigcup_{\xi \in (t-2,t)} D^{-}u(\xi) = \chi_{(3,6) \vee (6,8) \vee (9,11)}(t) \qquad (3)$$

and in (1), (2), (3) $t \in \boldsymbol{R}$. We infer that:

$t < 0$ $\quad x(t) = 0 \qquad (4)$

because $x \in Real$. For $t \geq 0$ we start using 5.6 (1) and we get:

| | | |
|---|---|---|
| $t \in [0,3)$ | $D^{-}x(t) = (0 \oplus 0) \cdot 1 \cdot \chi_{[2,\infty)}(t) \oplus 0 = 0$ | (5) |
| | $x(t) = 0$ | (6) |
| $t = 3$ | $D^{-}x(3) = (0 \oplus 0) \cdot 1 \cdot 1 \oplus 0 = 0$ | (7) |
| | $x(3) = 0$ | (8) |
| $t \in (3,6)$ | $D^{-}x(t) = (0 \oplus u(t-0)) \cdot 0 \cdot 1 \oplus 0 = 0$ | (9) |
| | $x(t) = 0$ | (10) |
| $t = 6$ | $D^{-}x(6) = (0 \oplus 0) \cdot 1 \cdot 1 \oplus 0 = 0$ | (11) |
| | $x(6) = 0$ | (12) |
| $t \in (6,8)$ | $D^{-}x(t) = (0 \oplus 1) \cdot 0 \cdot 1 \oplus 0 = 0$ | (13) |
| | $x(t) = 0$ | (14) |
| $t = 8$ | $D^{-}x(8) = (0 \oplus 1) \cdot 1 \cdot 1 \oplus 0 = 1$ | (15) |
| | $x(8) = 1$ | (16) |
| $t \in (8,9]$ | $D^{-}x(t) = (1 \oplus 1) \cdot 1 \cdot 1 \oplus 0 = 0$ | (17) |
| | $x(t) = 1$ | (18) |
| $t \in (9,11)$ | $D^{-}x(t) = (1 \oplus 0) \cdot 0 \cdot 1 \oplus 0 = 0$ | (19) |
| | $x(t) = 1$ | (20) |

$t = 11$ $D^- x(11) = (1 \oplus 0) \cdot 1 \cdot 1 \oplus 0 = 1$ (21)

$x(11) = 0$ (22)

$t > 11$ $D^- x(t) = (0 \oplus 0) \cdot 1 \cdot 1 \oplus 0 = 0$ (23)

$x(t) = 0$ (24)

i.e. the solution of 5.6 (1) is:

$$x(t) = \chi_{[8,11)}(t) \quad (25)$$

In this example, we have supposed that $u$ is equal with 0 everywhere except two time intervals: [3,4) and [6,9), the first of length less than $\tau$, the second of length $\geq \tau$. The conclusion is concordant with 5.5 in the sense that:

a) the first perturbation, corresponding to the interval [3,4), was filtered out

b) the second perturbation, corresponding to the interval [6,9), was shifted in time with $\tau$ time units.

5.8 **Definition** The delay model of the delay element and the delay itself are said to be:

a) *fixed*, if $\tau > 0$ is fixed (cf. [Davis, Nowick, 1997], [Lam, 1993])

b) *bounded*, if $\tau \in [m, M]$ is parameter and $m, M$ are fixed (cf. [Lavagno, 1992], [Davis, Nowick, 1997], [Lam, 1993])

c) *unbounded*, if $\tau > 0$ is parameter (cf. [Lavagno, 1992], [Davis, Nowick, 1997])

c') *unbounded*, if $\tau \in [0, M]$, $M$ fixed (cf. [Lam, 1993])

c") *unbounded*, if $\tau \in (0, M]$, with $M$ fixed

5.9 **Remark** The motivation of the classification 5.8 is given by the fact that $\tau$ may be not known and constant. It can depend on the technology, on the temperature and on the switching sense, from *LOW-HIGH*, respectively from *HIGH-LOW*. The bounds $m, M$ may refer to "all possible chips, including even those fabricated on different wafers", [Lam, 1993].

5.10 **Remark** We mention [Beerel, Meng, 1991] where the separation pure delay model-inertial delay model is made under the next form:

a) *transparent delay model*, informally defined by : "...allows propagation of short excitation pulses through gates and allows time varying delay elements"

b) *inertia delay model*, coinciding with the previous points of view.

In [Beerel, 1994] the *pure chaos delay element* pcde is informally defined by: "the function of a pcde is like a queue of transitions in which a sequence of transitions can accumulate before the first transition of the sequence emerges at the pcde's output". "When a new transition is pushed into a non-empty pcde, it may combine with the adjacent transition and the two transitions may annihilate each other". "It models the inertia quality of real gates in that real gates cannot react to very short excitation pulses. In a pcde, the choice between passing through and annihilating transitions is random, hence its chaotic nature". "For those transitions that do pass through the pcde, the delay is arbitrary, but finite".

The ideas that were presented at 5.9 allow that the delay is a function of time, generally unknown (perhaps it is bounded and the bounds are known). In fact, the transparent delay model and pcde are the pure delay model and the inertial delay model, our definitions 5.4 and 5.6, when $\tau = \tau(t)$.

5.11 **Definition** We have the next classification of the delays:

a) *wire delay model*:

$$x(t)=u(t-\tau) \quad (1)$$

where $u,x\in Real$ are the models of the electrical signals from the two ends of the wire

b) *gate delay model*; we note with $f:\mathbf{B}_2^m\to\mathbf{B}_2$ the function that is implemented by the gate, with $u_1,...,u_m,x\in Real$ the models of the inputs and of the state (of the output), with $\tau_1,...,\tau_m,\tau>0$ the delays and with $x^0\in\mathbf{B}_2$ the initial state and we have:

b.1) $$x(t)=f(u_1(t-\tau_1),...,u_m(t-\tau_m))\cdot\chi_{[0,\infty)}(t) \quad (2)$$

b.2) $$x(t)=f(u_1(t-\tau),...,u_m(t-\tau))\cdot\chi_{[\tau,\infty)}(t)\oplus x^0\cdot\chi_{[0,\tau)}(t) \quad (3)$$

b.3) $$D^-x(t)=(x(t-0)\oplus f(u_1(t-0),...,u_m(t-0)))\cdot \quad (4)$$

$$\cdot\overline{\bigcup_{\xi\in(t-\tau,t)}D^-f(u_1(\xi),...,u_m(\xi))}\cdot\chi_{[\tau,\infty)}(t)\oplus x^0\cdot\chi_{\{0\}}(t)$$

These equations give

b.1) the *input delay model*

b.2), b.3) the *output delay models*, the *pure*, respectively the *inertial* one.

5.12 **Informal definition** The previous definition appears under some informal form in all the cited works. Here is [Beerel, Meng, 1991]:

a) *input delay model*: "we assume instantaneous evaluation of the gate function with an independent delay element at each input lead of the gate"

b) *output delay model*: "we assume instantaneous evaluation of the gate function with an independent delay element at each output lead of the gate". Furthermore, a common form of output delay model is the *inertia delay model*, where "the output delay element is fixed and the gate does not react to excitation pulses shorter than the output delay".

5.13 **Remark** We give intuition on the definition 5.11.

First of all, as the asynchronous circuits consist in wires and gates, their delays appear on wires and gates.

Second, inertia is given (see 3.15) by the migration of a cloud of electrons inside a semi-conductor. The wire - implementing the identical Boolean function $\mathbf{B}_2\to\mathbf{B}_2$ - is a conductor with no inertial properties and the wire delay model is the pure delay model.

Third, the logical gates that implement the $\mathbf{B}_2^m\to\mathbf{B}_2$ functions can have input delays (at b.1)), or output delays (at b.2), b.3)), the lack of delays being a non-interesting idealization in the present paper and the presence of the delays on inputs and outputs being reduced at b.1), b.2), b.3). The case b.1) shows how $f$ can be computed when its inputs, for various reasons, arrive with delays. As the computation of $f$ only, made with a semi-conductor, can be made inertially and the input delay model is not related to the computation of $f$, b.1) is a pure delay model (there are $m$ pure delay models there). The situation differs when we refer in b.2), b.3) to the computation of $f$: it can be approximated in two ways, the non-inertial and the inertial one.

5.14 **Remark** 5.11 b.3) can be analyzed by means of an example similar to 5.7, where we replace $u$ with $f\circ u$.

5.15 **Remark** The delays on wires and gates should be understood as primitive, atomic delays of the asynchronous circuits. In fact, in [Davis, Nowick, 1997] it is given the next classification of the delay models:

- *simple gate* (*gate level*) *models*
- *complex gate models*, where "an entire sub-network of gates is modeled by a single delay".

5.16 **Example** of complex gate model. Let the next circuit:

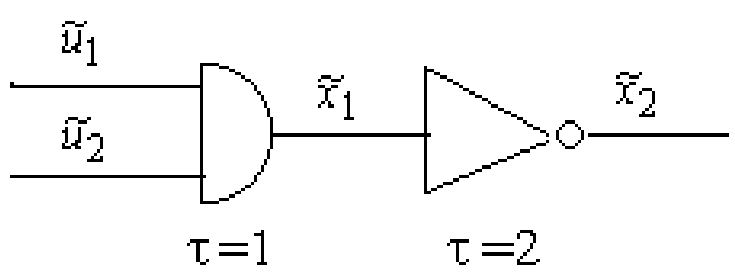

fig (1)

described in the output pure delay model by the equations:

$$x_1(t) = u_1(t-1) \cdot u_2(t-1) \cdot \chi_{[1,\infty)}(t) \oplus x_1^0 \cdot \chi_{[0,1)}(t) \quad (1)$$

$$x_2(t) = \overline{x_1(t-2)} \cdot \chi_{[2,\infty)}(t) \oplus x_2^0 \cdot \chi_{[0,2)}(t) \quad (2)$$

We get:

$$x_2(t) = \overline{u_1(t-3) \cdot u_2(t-3) \cdot \chi_{[1,\infty)}(t-2) \oplus x_1^0 \cdot \chi_{[0,1)}(t-2)} \cdot \chi_{[2,\infty)}(t) \oplus x_2^0 \cdot \chi_{[0,2)}(t) \quad (3)$$

$$= \overline{u_1(t-3) \cdot u_2(t-3) \cdot \chi_{[3,\infty)}(t) \oplus x_1^0 \cdot \chi_{[2,3)}(t)} \cdot \chi_{[2,\infty)}(t) \oplus x_2^0 \cdot \chi_{[0,2)}(t)$$

$$= (\overline{u_1(t-3) \cdot u_2(t-3)} \cdot \chi_{[3,\infty)}(t) \oplus \overline{x_1^0} \cdot \chi_{[2,3)}(t)) \cdot \chi_{[2,\infty)}(t) \oplus x_2^0 \cdot \chi_{[0,2)}(t)$$

If

$$\overline{x_1^0} = x_2^0 \overset{not}{=} x^0 \quad (4)$$

we have

$$x_2(t) = \overline{u_1(t-3) \cdot u_2(t-3)} \cdot \chi_{[3,\infty)}(t) \oplus x^0 \cdot \chi_{[0,3)}(t) \quad (5)$$

i.e. instead of two gates like in fig (1), we can take in consideration a circuit that

- implements the composed Boolean function
- has the delay given by the sum 3=1+2.

5.17 **Remark** In the previous example, the supposition (4) does not alter the quality of the model. In the general case however, a discussion is necessary on the values of the initial states.

On the other hand, the possibility of using two "simple gate models" giving a "complex gate model" is a consequence of the fact that both "simple gate models" are pure.

5.18 **Remark** If both gate models from 5.16 (1) are inertial, the complex gate model of the circuit has a delay equal with 3=1+2, but it filters out pulses shorter than 2, instead of filtering out pulses shorter than 3. The general rule here is: the delay is given by $\tau_1 + ... + \tau_n$ and filtering out refers to $\max\{\tau_1, ..., \tau_n\}$.

5.19 **Remark** One of the meanings of the inertial delay with $\tau$ time units 5.11 (4) is the following: if $t \geq \tau$ and

$$D^{-}x(t)=1 \qquad (1)$$

then

$$\bigcup_{\xi\in(t-\tau,t)} D^{-} f(u_1(\xi),...,u_m(\xi))=0 \qquad (2)$$

thus $f(u_1,...,u_m)_{|[t-\tau,t)}$ is the constant function, necessary condition for the switch (1) to happen.

We have that for the inertial complex gate model of the asynchronous circuit from fig 5.16 (1), $(u_1 \cdot u_2)_{|[t-3,t)}$ constant is too strong compared to $(u_1 \cdot u_2)_{|[t-2,t)}$ constant, necessary condition for $x_2$ to switch.

We accept that the inertial delay model and as a special case the complex gate inertial delay model uses data in conditions probably stronger than necessary. From this point of view, modeling 5.16 (1) by:

$$D^{-}x_2(t)=(x_2(t-0)\oplus \overline{u_1(t-0)\cdot u_2(t-0)})\cdot \overline{\bigcup_{\xi\in(t-3,t)} D^{-}\overline{u_1(\xi)\cdot u_2(\xi)}}\cdot \chi_{[3,\infty)}(t)\oplus x_2^0\cdot \chi_{\{0\}}(t) \quad (3)$$

is reasonable.

5.20 **Remark** Let us quote [Liebelt, 1995] saying "A common form of the implementation of the inertial delay model is the one in which the transmission delay for the transitions is the same with the threshold for cancellation".

The author accepts two parameters for the inertial delay model - that we have seen to govern the complex gate inertial delay model from 5.16 - that he calls *transmission delay* and *threshold for cancellation* and he mentions that usually they are assumed to be equal. We give the equation characterizing the case when the two delays differ:

$$D^{-}x(t)=(x(t-0)\oplus f(u_1(t-\delta-0),...,u_m(t-\delta-0)))\cdot \qquad (4)$$

$$\cdot \overline{\bigcup_{\xi\in(t-\tau-\delta,t-\delta)} D^{-} f(u_1(\xi),...,u_m(\xi))}\cdot \chi_{[\tau+\delta,\infty)}(t)\oplus x^0\cdot \chi_{\{0\}}(t)$$

where $\tau>0,\delta\geq 0$, $\tau$ is the cancellation delay, $\tau+\delta$ is the transmission delay and $\tau\leq\tau+\delta$, like at 5.16.

We shall not use this model.

5.21 **Remark** Another source of approximation is given by the values that the model function $x$ has during the switching interval $[t_1,t_2]$ of $\tilde{x}$, see fig 3.1 (1). We have adopted the binary logic for algebraical reasons ( $\boldsymbol{B}_2$ is a field, but a three element set cannot be organized similarly) and it is not clear so far what values $x$ must have when $\tilde{x}$ crosses the *UNCERTAINTY* range of values. We need the next

5.22 **Convention** We ask referring to fig 3.1 (1) again that

$$x_{|[t_1,t_2]}=x(t_1-0) \qquad (1)$$

i.e. during the switch, the model keeps its previous value.

5.23 **Remark** The condition 5.22 (1) is fulfilled if we take $\tau$ big enough (in its own range of acceptable values) so that when the switch of $x$ happens i.e. when 5.19 (1) is true, the uncertainties are already ended. We read 5.19 (1) like this: " $x$ has surely switched at $t$ " (or previously).

**6. Asynchronous Automata**

6.1 **Notation** We note with $\{\emptyset\}$ the set formed by one single element.

6.2 **Remark** For the sake of a unitary exposure, we give $\{\emptyset\}$ the next meaning: if the argument of a function has a variable running in $\{\emptyset\}$, then the function does not depend on that variable; and if a function has the values in the set $\{\emptyset\}$, then it is constant (it takes one value).

6.3 **Definition** We call *asynchronous automaton*, or *asynchronous system*, a mathematical concept $\Sigma$ given by the following data:

- the *time set* is $\boldsymbol{R}$
- $t \in \boldsymbol{R}$ is the free variable, called the *time moment*, or the *time instant*
- $t_0 = 0 \in \boldsymbol{R}$ is called the *initial time* (*moment*, or *instant*)
- $n \in \boldsymbol{N}$ is called the *dimension of the state space*, or the *dimension of* $\Sigma$
- the *state space* $X$ is given by

$$X = \begin{cases} \{\emptyset\}, n = 0 \\ \boldsymbol{B}_2^n, n \geq 1 \end{cases} \tag{1}$$

- $x \in Real^{(n)}$ is called the *state*, or the *state function*, or the *trajectory* (*of* $\Sigma$). $x$ is also called *solution* (of an equation), *orbit* or *field line*. Sometimes we speak about the *states*, or the *state variables* of the automaton and this refers to the coordinate functions $x_1,...,x_n$. The first $n_1 \in \{0,...,n\}$ of them are called *ideal* (or *pure*, or *non-inertial*) *states* (or *coordinates*) and the last $n - n_1$ of them are called *inertial states* (or *coordinates*). We consider that $x \in Real^{(0)}$ represents the null state function.

- $x^0 \in X$ is called the *initial state*
- $m \in \boldsymbol{N}$ is called the *dimension of the input space*
- $U$ is called the *input space*

$$U = \begin{cases} \{\emptyset\}, m = 0 \\ \boldsymbol{B}_2^m, m \geq 1 \end{cases} \tag{2}$$

- $u \in Real^{(m)}$ is called the *input*, or the *input function*, or the *control* (*of* $\Sigma$); sometimes we speak about the *inputs* of the automaton and this refers to the coordinate functions $u_1,...,u_m$. We consider that $u \in Real^{(0)}$ represents the null input function.

- the function $f : \boldsymbol{B}_2^n \times \boldsymbol{B}_2^m \to \boldsymbol{B}_2^n$ is called the *generator function* (*of* $\Sigma$). If $m = 0$ and $n \geq 1$, we can consider that it is given a generator function noted $g : \boldsymbol{B}_2^n \to \boldsymbol{B}_2^n$ and if $n = 0$ - whichever $m$ might be- we can consider that no generator function is given

- the positive parameters $\tau_i, i = \overline{1,n}$ are called the *delay parameters*, or the *switching time parameters* (*of the coordinate functions* $f_1,..., f_n$ of $f$, respectively *of the coordinate functions* $g_1,...,g_n$ of $g$)

- let us suppose that $n \geq 1$; then the relation between the previous data is given by:

case $m \geq 1$: the equation

$$x_i(t) = f_i(x(t-\tau_i), u(t-\tau_i)) \cdot \chi_{[\tau_i,\infty)}(t) \oplus x_i^0 \cdot \chi_{[0,\tau_i)}(t), i = \overline{1, n_1} \tag{3}$$

$$D^- x_i(t) = (x_i(t-0) \oplus f_i(x(t-0), u(t-0))) \cdot \tag{4}$$

$$\cdot \overline{\bigcup_{\xi \in (t-\tau_i, t)} D^- f_i(x(\xi), u(\xi)) \cdot \chi_{[\tau_i,\infty)}(t) \oplus x_i^0 \cdot \chi_{\{0\}}(t)}, i = \overline{n_1 + 1, n}$$

case $m = 0$: the equation

$$x_i(t) = g_i(x(t-\tau_i)) \cdot \chi_{[\tau_i,\infty)}(t) \oplus x_i^0 \cdot \chi_{[0,\tau_i)}(t), i = \overline{1, n_1} \tag{5}$$

$$D^- x_i(t) = (x_i(t-0) \oplus g_i(x(t-0))) \cdot \tag{6}$$

$$\cdot \overline{\bigcup_{\xi \in (t-\tau_i, t)} D^- g_i(x(\xi)) \cdot \chi_{[\tau_i,\infty)}(t) \oplus x_i^0 \cdot \chi_{\{0\}}(t)}, i = \overline{n_1 + 1, n}$$

In (3),...,(6) one of the sets $\{1,...,n_1\}, \{n_1+1,...,n\}$ can be empty and in this situation that equation is missing. In these equations $t \in \boldsymbol{R}$; $x^0$ and, in case that it exists, $u$ are given, $\tau_1,...,\tau_n$ are parameters and $x$ is the unknown.

- if $n = 0$, then none of (3),...,(6) exists.

6.4 **Remark** We consider that the words *automaton* and *system* are synonyms and they express the abstractization of the notion of *circuit*, as we have already said at 3.3. Sometimes, by the word system we shall refer to one of the systems of equation 6.3 (3), (4), respectively 6.3 (5), (6) but this will create no confusions.

6.5 **Remark** By *asynchronous* system, in the systems theory we usually understand real time systems, where the variables run in $\boldsymbol{B}_2^k$ spaces.

6.6 **Definition** An automaton for which $m = 0$ is called *autonomous*; if $m > 0$, the automaton is called *non-autonomous*, or *controlled*, or *control automaton* and an automaton for which $n = 0$ is called *trivial*. If $m = n = 0$, then the automaton is called *empty*, or *void*.

6.7 **Remark** A non-empty trivial automaton consists in the input $u$ and no relation (of determinism). The empty automata are the automata with no content.

6.8 **Definition** The equations 6.3 (3), (4) are called the *equations of the* (*non-autonomous*, or *controlled*) *asynchronous automata*, shortly EAA. The equations 6.3 (5), (6) are called the *equations of the autonomous asynchronous automata*, shortly EAAA.

6.9 **Remark** There exists a relation between the autonomous and the controlled automata, given by the situation when $m \geq 1$ and in EAA $u$ is the constant function:

$$u(t) = u^0 \cdot \chi_{[0,\infty)}(t), u^0 \in \boldsymbol{B}_2^m \tag{1}$$

Then the automaton $\Sigma$ described by 6.3 (3), (4) behaves like the automaton $\Sigma'$ described by 6.3 (5), (6), the next equation being true:

$$g(\cdot) = f(\cdot, u^0) \tag{2}$$

6.10 **Definition** The previous automaton $\Sigma$ is called *autonomous-like*.

6.11 **Remark** The supposition that EAA, respectively EAAA contain both types of coordinates, non-inertial and inertial does not restrict the generality. This fact happens because if $x_i^0 = 0$ and $f_i = 0$, respectively $g_i = 0$, then $x_i = 0$ is a solution. Thus, in the situation when pure or inertial coordinates do not exist, a null coordinate can be added.

On the other hand, if EAA, respectively EAAA contain coordinates where $x_i^0 = 0$ and $f_i = 0$, respectively $g_i = 0$, the equations can be rewritten eliminating these coordinates.

6.12 **Definition** In the special case when in EAA or in EAAA the generator function does not depend on *x*, $\Sigma$ is called *combinational automaton*, or *automaton without feedback* and otherwise $\Sigma$ is called *sequential*, or *with feedback*.

6.13 **Definition** Suppose that the Boolean function that is implemented by a circuit satisfies the property that it depends only on the input; then the circuit is called *combinational* (*without feedback*) itself. In the case when the function depends on the state (or output) also, the asynchronous circuit is called *sequential* (*with feedback*)

6.14 **Remark** EAA and EAAA give an output delay model of the asynchronous circuits characterized by:

- a combination of the pure delay model and the inertial delay model
- a combination of the simple gate model and the complex gate model
- any of the fixed, bounded and unbounded delay models from 5.8.

These equations do not match exactly the transparent delay model and the pure chaos delay model of Beerel (see 5.10), but in the paragraph 19 of our work related to the branching time temporal logic we shall make some comments on this topic too.

6.15 **Remark** We give the dual, anticipative version of EAA

$$x_i(t) = f_i(x(t+\tau_i), u(t+\tau_i)) \cdot \chi_{(-\infty,-\tau_i]}(t) \oplus x_i^0 \cdot \chi_{(-\tau_i,0]}(t),\ i = \overline{1, n_1} \qquad (1)$$

$$D^+ x_i(t) = (x_i(t+0) \oplus f_i(x(t+0), u(t+0))) \cdot \qquad (2)$$

$$\cdot \overline{\bigcup_{\xi \in (t, t+\tau_i)} D^+ f_i(x(\xi), u(\xi))} \cdot \chi_{(-\infty,-\tau_i]}(t) \oplus x_i^0 \cdot \chi_{\{0\}}(t),\ i = \overline{n_1 + 1, n}$$

$u \in Real^{*(m)}, x \in Real^{*(n)}$ and the rest of the data remain the same.

The solution $x'$ of (1), (2) and the solution $x$ of EAA, supposing that such solutions exist and are unique (to be proved later) run the time set in opposite senses:

$$x(t) = x'(-t), t \in \mathbf{R} \qquad (3)$$

6.16 **Remark** There exist other similar ways of writing the equations of the asynchronous automata, that bring nothing essentially new. It is possible to initialize the coordinates $x_1, ..., x_n$ not starting with the time instant $t_0 = 0$ but ending with the time instant $t_1$.

6.17 **Remark** In [Kalman+, 1975], when defining the (finite) automaton it is asked, unlike we do, that the time set is discrete: $\mathbf{Z}$ or $\mathbf{N}$. The existing relation between the real and the discrete time will be one of our concerns and as we shall see the two points of view, ours and theirs, do not differ.

## 7. Example: The Clock Generator

7.1 **Example** of autonomous asynchronous automaton. The next circuit:

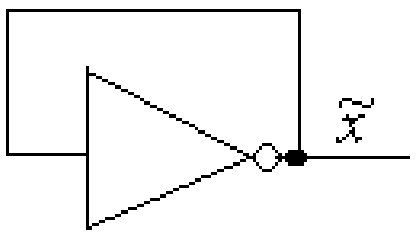

fig (1)

is called the *clock generator*. In the non-inertial, respectively in the inertial variant, using the unbounded delay model 5.8 c'') EAAA that model this circuit are, with the generator function $g:\boldsymbol{B}_2\rightarrow\boldsymbol{B}_2$ = the logical complement and after simple computations:

$$x(t)=\overline{x(t-\tau)}\cdot\chi_{[\tau,\infty)}(t)\oplus x^0\cdot\chi_{[0,\tau)}(t) \qquad (2)$$

$$D^{-}x(t)=\overline{\bigcup_{\xi\in(t-\tau,t)}D^{-}x(\xi)}\cdot\chi_{[\tau,\infty)}(t)\oplus x^0\cdot\chi_{\{0\}}(t) \qquad (3)$$

with $x^0\in\boldsymbol{B}_2, x\in Real$ and $\tau\in(0,M]$.

7.2 **Theorem** The solutions of 7.1 (2), (3) are unique, they coincide and they are given by:

$$x(t)=\begin{cases}0 & ,t<0\\ x^0 & ,t\in\bigvee_{k\in\boldsymbol{N}}[2k\tau,(2k+1)\tau)\\ 1\oplus x^0 & ,t\in\bigvee_{k\in\boldsymbol{N}}[(2k+1)\tau,(2k+2)\tau)\end{cases}=x^0\cdot\chi_{[0,\infty)}(t)\oplus\Xi_{k\geq1}\chi_{[k\tau,\infty)}(t) \qquad (1)$$

**Proof** 7.1 (2) is solved easily, taking into account that:

$$x(t)=\begin{cases}0 & ,t<0 \text{ because } x \text{ is realizable}\\ x^0 & ,t\in[0,\tau) \text{ from 7.1 (2)}\end{cases} \qquad (2)$$

and considering then $t\in[\tau,2\tau), t\in[2\tau,3\tau),...$

We solve 7.1 (3):

| | | |
|---|---|---|
| $t<0$, | $x(t)=0$ | (3) |
| $t=0$, | $D^{-}x(0)=x^0$ | (4) |
| | $x(0)=x^0$ | (5) |
| $t\in(0,\tau)$, | $D^{-}x(t)=0$ | (6) |
| | $x(t)=x^0$ | (7) |
| $t=\tau$, | $D^{-}x(\tau)=1$ | (8) |
| | $x(\tau)=1\oplus x^0$ | (9) |
| $t\in(\tau,2\tau)$, | $D^{-}x(t)=0$ | (10) |
| | $x(t)=1\oplus x^0$ | (11) |
| $t=2\tau$, | $D^{-}x(2\tau)=1$ | (12) |
| | $x(2\tau)=1\oplus(1\oplus x^0)=x^0$ | (13) |

$t \in (2\tau, 3\tau)$, $\quad D^- x(t) = 0$ (14)

$x(t) = x^0$ (15)

...

7.3 **Remark** There exists an ambiguity in the study of this automaton given by the next property, see 7.2 (1):

$$\forall t \in (0,\infty), \exists \tau_1, \tau_2 \in (0, M], \exists k_1, k_2 \in \boldsymbol{N},$$
$$t \in [2k_1\tau_1, (2k_1+1)\tau_1) \wedge [(2k_2+1)\tau_2, (2k_2+2)\tau_2)$$

This means that for no time instant $t > 0$, we can say whether $x$ models the electrical signal $\tilde{x}$ of the clock generator.

The fulfillment of the previous property becomes obvious if we take

$$\tau_1 \in (\frac{t}{2k_1+1}, \frac{t}{2k_1}] \wedge (0, M], \ \tau_2 \in (\frac{t}{2k_2+2}, \frac{t}{2k_2+1}] \wedge (0, M]$$

and this is possible if, for some $t$, we choose $k_1, k_2 \geq 1$ sufficiently great so that the meets are non-empty.

7.4 **Remark** It is not clear so far how do we characterize the circuits that accept a model function $x$ of the electrical signal $\tilde{x}$, at least on an non-empty set $I \subset [0,\infty)$, or perhaps on all of $[0,\infty)$, so that $x$ is the solution of EAA or of EAAA.

7.5 **Remark** Even in this situation when modeling is not possible, the *form* of $x(t)$ from 7.2 (1), switching on and on from 0 to 1 and from 1 to 0 resembles the *form* of the electrical signal $\tilde{x}$ of the clock generator, as $\tilde{x}$ switches on and on from *LOW-HIGH* and from *HIGH-LOW.*

7.6 **Remark** The name clock generator of the circuit from 7.1 (1) is justified by the fact that $\tilde{x}$ counts the discrete time when switching permanently from *LOW-HIGH* and from *HIGH-LOW.*

**8. Example: The R-S Latch**

8.1 **Example** of autonomous-like asynchronous automaton. The next drawing

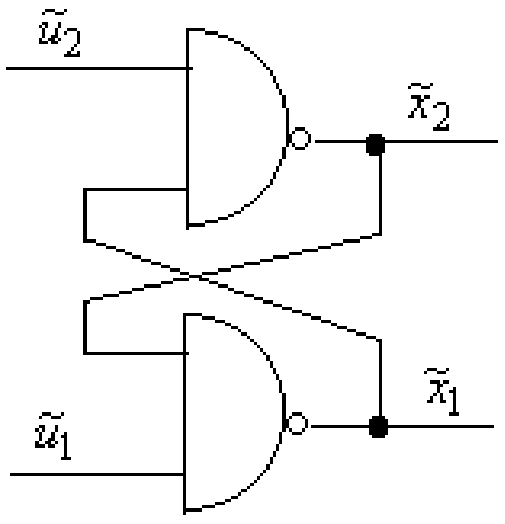

fig (1)

refers to the asynchronous circuit called *R-S latch*. We shall suppose that the inputs are:

$$u_1(t) = R \cdot \chi_{[0,\infty)}(t) \quad (2)$$

$$u_2(t) = S \cdot \chi_{[0,\infty)}(t) \quad (3)$$

$R, S \in \boldsymbol{B}_2$ and the following equations result after some elementary computations

$$x_1(t) = \overline{R \cdot x_2(t-\tau_1)} \cdot \chi_{[\tau_1,\infty)}(t) \oplus x_1^0 \cdot \chi_{[0,\tau_1)}(t) \quad (4)$$

$$x_2(t) = \overline{S \cdot x_1(t-\tau_2)} \cdot \chi_{[\tau_2,\infty)}(t) \oplus x_2^0 \cdot \chi_{[0,\tau_2)}(t) \quad (5)$$

respectively

$$D^{-}x_1(t)=(x_1(t-0)\oplus \overline{R\cdot x_2(t-0)})\cdot R\cdot \overline{\bigcup_{\xi\in(t-\tau_1,t)} D^{-}x_2(\xi)}\cdot\chi_{[\tau_1,\infty)}(t)\oplus x_1^0\cdot\chi_{\{0\}}(t) \quad (6)$$

$$D^{-}x_2(t)=(x_2(t-0)\oplus \overline{S\cdot x_1(t-0)})\cdot S\cdot \overline{\bigcup_{\xi\in(t-\tau_2,t)} D^{-}x_1(\xi)}\cdot\chi_{[\tau_2,\infty)}(t)\oplus x_2^0\cdot\chi_{\{0\}}(t) \quad (7)$$

where $x^0\in \boldsymbol{B}_2^2, x\in Real^{(2)}, \tau_1\in(0,M_1]$ and $\tau_2\in(0,M_2]$. We have used once again the unbounded delay model 5.8 c").

To be solved the two systems of equations: (2), (3), (4), (5), respectively (2), (3), (6), (7) by taking in consideration the next possibilities:

a) $\tau_1<\tau_2$ b) $\tau_1=\tau_2$ c) $\tau_1>\tau_2$

i) $R=0,S=0$ ii) $R=1,S=0$ iii) $R=0,S=1$ iv) $R=1,S=1$

j) $x_1^0=x_2^0$ jj) $x_1^0\neq x_2^0$

8.2 **Theorem** By noting with $x\in Real^{(2)}$ the solution of 8.1 (2), (3), (4), (5) and with $y\in Real^{(2)}$ the solution of 8.1 (2), (3), (6), (7), we have that $x, y$ are unique and they are given, with a),…,c), i),…,iv), j),…, jj) like at 8.1, by:

i)
$$\begin{cases} x_1(t)=y_1(t)=x_1^0\cdot\chi_{[0,\tau_1)}(t)\oplus\overline{R}\cdot\chi_{[\tau_1,\infty)}(t)\\ x_2(t)=y_2(t)=x_2^0\cdot\chi_{[0,\tau_2)}(t)\oplus\overline{S}\cdot\chi_{[\tau_2,\infty)}(t)\end{cases} \quad (1)$$

ii) For a), b):

$$\begin{cases} x_1(t)=y_1(t)=x_1^0\cdot\chi_{[0,\tau_1)}(t)\oplus\overline{x_2^0}\cdot\chi_{[\tau_1,\tau_1+\tau_2)}(t)\oplus\overline{R}\cdot\chi_{[\tau_1+\tau_2,\infty)}(t)\\ x_2(t)=y_2(t)=x_2^0\cdot\chi_{[0,\tau_2)}(t)\oplus\overline{S}\cdot\chi_{[\tau_2,\infty)}(t)\end{cases} \quad (2)$$

In the case c), $x_1, x_2, y_2$ coincide with these from (2), while $y_1$ is given by:

- if j) is true, then

$$y_1(t)=x_1^0\cdot\chi_{[0,\tau_1)}(t)\oplus\overline{R}\cdot\chi_{[\tau_1,\infty)}(t) \quad (3)$$

- if jj) is true, then

$$y_1(t)=x_1^0\cdot\chi_{[0,\tau_1+\tau_2)}(t)\oplus\overline{R}\cdot\chi_{[\tau_1+\tau_2,\infty)}(t) \quad (4)$$

iii) is similar with ii)

iv)
$$\begin{cases} x_1(t)=\underset{k\in\boldsymbol{N}}{\Xi}\, x_1^0\cdot\chi_{[k(\tau_1+\tau_2),k(\tau_1+\tau_2)+\tau_1)}(t)\oplus\underset{k\in\boldsymbol{N}}{\Xi}\,\overline{x_2^0}\cdot\chi_{[k(\tau_1+\tau_2)+\tau_1,(k+1)(\tau_1+\tau_2))}(t)\\ x_2(t)=\underset{k\in\boldsymbol{N}}{\Xi}\, x_2^0\cdot\chi_{[k(\tau_1+\tau_2),k(\tau_1+\tau_2)+\tau_2)}(t)\oplus\underset{k\in\boldsymbol{N}}{\Xi}\,\overline{x_1^0}\cdot\chi_{[k(\tau_1+\tau_2)+\tau_2,(k+1)(\tau_1+\tau_2))}(t)\end{cases} \quad (5)$$

If a) is true, then

$$\begin{cases} y_1(t)=x_1^0\cdot\chi_{[0,\tau_1)}(t)\oplus\overline{x_2^0}\cdot\chi_{[\tau_1,\infty)}(t)\\ y_2(t)=x_2^0\cdot\chi_{[0,\infty)}(t)\end{cases} \quad (6)$$

If b) is satisfied, $y_1, y_2$ coincide with $x_1, x_2$ from (5) and

If c) is true:

$$\begin{cases} y_1(t) = x_1^0 \cdot \chi_{[0,\infty)}(t) \\ y_2(t) = x_2^0 \cdot \chi_{[0,\tau_2)}(t) \oplus \overline{x_1^0} \cdot \chi_{[\tau_2,\infty)}(t) \end{cases} \tag{7}$$

8.3 **Remark** If

$$R \cdot S = 0 \tag{1}$$

then the next implications are true for both variants, pure delay and inertial delay model:

$$t \geq \tau_1 + \tau_2 \Rightarrow x_1(t) = y_1(t) = \overline{R} \tag{2}$$

$$t \geq \tau_1 + \tau_2 \Rightarrow x_2(t) = y_2(t) = \overline{S} \tag{3}$$

These equations give the conclusion, see the definition 3.10, that $x_1, x_2$ and $y_1, y_2$ model $\tilde{x}_1, \tilde{x}_2$ on $[M_1 + M_2, \infty)$. To be compared with the situation from paragraph 7.

When (1) is not valid, (2) and (3) are not valid themselves. Then the possibility of having a model satisfying EAA at the same time, does not exist.

8.4 **Remark** The circuit from 8.1 (1) together with the inputs 8.1 (2), (3), if 8.3 (1) is true and taking into account the implications 8.3 (2), (3) - give the next complex gate model for the R-S latch in the pure delay case:

$$\begin{cases} x_1(t) = x_1^0 \cdot \chi_{[0,\tau_1+\tau_2)}(t) \oplus \overline{R} \cdot \chi_{[\tau_1+\tau_2,\infty)}(t) \\ x_2(t) = x_2^0 \cdot \chi_{[0,\tau_1+\tau_2)}(t) \oplus \overline{S} \cdot \chi_{[\tau_1+\tau_2,\infty)}(t) \end{cases} \tag{1}$$

The advantages and the shortcomings of the model are obvious.

## 9. The Modeling of the Asynchronous Circuits

9.1 **Remark** In the paragraph 3 we have seen how a realizable function $x$ models an electrical signal $\tilde{x}$ (on a set $I \subset [0,\infty)$) and this was always possible, cf. with Proposition 3.18. The examples from 7, 8 have shown however a different, systemic way of putting the modeling problem, i.e. referring to an asynchronous circuit (and to a set $I \subset [0,\infty)$) and this is not always possible. Even if, in the last case, modeling is possible on some set *I*, it may happen that *I* is not known.

9.2 **The purpose of this section** is that of discussing the modeling of the asynchronous circuits. Because their definition is informal - as it is not very clear what is it a 'wire', respectively a 'logic gate', all this section is informal.

9.3 **Informal definition** There are given EAA and the asynchronous circuit $\tilde{\Sigma}$. We say that EAA *is associated to* $\tilde{\Sigma}$ if

a) the inputs of $\tilde{\Sigma}$ are the signals $\tilde{u}_1,...,\tilde{u}_m$ and the inputs $u_1,...,u_m \in Real$ are their models (on $[0,\infty)$)

b) there is given a decomposition $\tilde{\Sigma}_1,...,\tilde{\Sigma}_n$ of $\tilde{\Sigma}$ in sub-circuits (the limit situation is when $\tilde{\Sigma}_1,...,\tilde{\Sigma}_n$ are all of them wires and logic gates), so that

- $\tilde{\Sigma}_1,...,\tilde{\Sigma}_n$ have exactly one output $\tilde{x}_1,...,\tilde{x}_n$, their inputs being some of $\tilde{u}_1,...,\tilde{u}_m, \tilde{x}_1,...,\tilde{x}_n$ (the limit situation is when for some $\tilde{\Sigma}_i$, none of them is input and then we say that $\tilde{x}_i$ is 'stuck at 0', or 'stuck at 1', i.e. it is constant from a logical point of view)

- $x_i^0 = \begin{cases} \nu(\tilde{x}_i(0)), if\ \tilde{x}_i(0) \in LOW \vee HIGH \\ arbitrary, else \end{cases}, i = \overline{1,n}$ where $\nu$ is the logical value function that was defined at 3.9

- $f_1,..., f_n : \boldsymbol{B}_2^n \times \boldsymbol{B}_2^m \to \boldsymbol{B}_2$ are the Boolean functions that $\tilde{\Sigma}_1,...,\tilde{\Sigma}_n$ implement

- $\tau_1,...,\tau_n$ are the delays that are introduced by $\tilde{\Sigma}_1,...,\tilde{\Sigma}_n$, the first $n_1$ of them being pure and the last $n - n_1$ of them being inertial.

9.4 **Remark** It is obvious the situation when, at 9.3, $\tilde{\Sigma}$ has no inputs.

9.5 **Informal definition** If we identify an automaton $\Sigma$ with a system of equations EAA or EAAA - this was suggested at 6.4 - then we say that $\Sigma$ *is associated to* $\tilde{\Sigma}$.

9.6 **Remark** Because being given $\tilde{\Sigma}$ and the sub-circuits $\tilde{\Sigma}_1,...,\tilde{\Sigma}_n$:

- there are several ways of choosing $u_1,...,u_m$ so that they model $\tilde{u}_1,...,\tilde{u}_m$

- there are several ways of giving arbitrary values to $x_i^0$ when $\nu(\tilde{x}_i(0)) \notin LOW \vee HIGH$, $i = \overline{1,n}$

- there are several ways of choosing $\tau_1,...,\tau_n$

we conclude that we have several EAA (EAAA) and several automata $\Sigma$ that are associated to $\tilde{\Sigma}$.

9.7 **Notation** We note with $[\tilde{\Sigma}]$ the set of the automata $\Sigma$ that are associated to $\tilde{\Sigma}$.

9.8 **Remark** The study of $\tilde{\Sigma}$ consists is characterizing the things that the automata $\Sigma \in [\tilde{\Sigma}]$ have in common. For example, in the case of the R-S latch from 8.1 (1) we know that $R \cdot S = 0$ implies

$$t \geq M_1 + M_2 \Rightarrow x_1(t) = \overline{R} \qquad (1)$$

$$t \geq M_1 + M_2 \Rightarrow x_2(t) = \overline{S} \qquad (2)$$

while the common property in the case of the clock generator from 7.1 (1) looks to be weaker, because, in the unbounded delay model, it is a qualitative and not a quantitative one:

$$\forall t, \exists t' > t, x(t) \neq x(t') \qquad (3)$$

9.9 **Informal definition** Consider the fixed delay model, when $\tau_1,...,\tau_n$ are known. If there exists a non-empty set $I \subset [0,\infty)$, depending on $\tau_1,...,\tau_n$, so that for any $\Sigma \in [\tilde{\Sigma}]$ $x_1,...,x_n$ model $\tilde{x}_1,...,\tilde{x}_n$ on $I$, we say that $\tilde{\Sigma}$ *accepts a model on* $I$ and in this situation any $\Sigma \in [\tilde{\Sigma}]$ is called a *model* of $\tilde{\Sigma}$ *on* $I$. This definition usually refers to the greatest set $I$ with the previous property and then $I$ itself must not be mentioned. We use to say that $\tilde{\Sigma}$ *accepts a model* and $\Sigma$ is a *model* of $\tilde{\Sigma}$.

9.10 **Remark** Similar definitions are given for the other delay models and here is an example. Let the unbounded delay model, where $\tau_1 \in (0, M_1],...,\tau_n \in (0, M_n]$ are parameters and $M_1,...,M_n$ are known. If there is a non-empty set $I \subset [0,\infty)$ depending on $M_1,...,M_n$, so that … from here the definition repeats 9.9.

9.11 **Remark** Let the unbounded delay model, version 5.8 c"). The essential difference between the clock generator and the R-S latch is now expressed by: the first does not accept a model, whilst the second accepts a model (on $[M_1 + M_2, \infty)$). The conditions in which a circuit accepts a model will be discussed in the next paragraphs (synchronous-likeness).

## 10. The Solutions of the Equations of the Asynchronous Automata

10.1 **Problem** To be solved the system of equations of the asynchronous automata EAA

$$x_i(t) = f_i(x(t-\tau_i), u(t-\tau_i)) \cdot \chi_{[\tau_i,\infty)}(t) \oplus x_i^0 \cdot \chi_{[0,\tau_i)}(t), i = \overline{1, n_1} \qquad (1)$$

$$D^- x_i(t) = (x_i(t-0) \oplus f_i(x(t-0), u(t-0))) \cdot \qquad (2)$$

$$\cdot \overline{\bigcup_{\xi\in(t-\tau_i,t)} D^- f_i(x(\xi), u(\xi)) \cdot \chi_{[\tau_i,\infty)}(t)} \oplus x_i^0 \cdot \chi_{\{0\}}(t), i = \overline{n_1 + 1, n}$$

with $f : \boldsymbol{B}_2^n \times \boldsymbol{B}_2^m \to \boldsymbol{B}_2^n$ and $t \in \boldsymbol{R}$; $x^0 \in \boldsymbol{B}_2^n$, $\tau_1, ..., \tau_n > 0$ and $u \in Real^{(m)}$,

$$u(t) = \underset{k\in \boldsymbol{N}}{\Xi}\, u^k \cdot \chi_{[\nu_k, \nu_{k+1})}(t) \qquad (3)$$

are given - where $u^k \in \boldsymbol{B}_2^m, k \in \boldsymbol{N}$ and the family

$$0 = \nu_0 < \nu_1 < \nu_2 < ... \qquad (4)$$

is SINLF, see 2.25; $x \in Real^{(n)}$ is the unknown.

We have supposed without loss that $1 \le n_1 < n$, see 6.11.

10.2 **Remark** The Problem 10.1 refers to the fixed delay model, cf. 5.8 a) and this is convenient for the present stage of analysis of EAA. Later we shall suppose that $\tau_1 \in (0, M_1], ..., \tau_n \in (0, M_n]$, the unbounded delay model, cf. 5.8 c").

The main result of this paragraph is expressed by Theorem 10.9. Now we shall need some preliminary results.

10.3 **Lemma** Let $z : \boldsymbol{R} \to \boldsymbol{B}_2$ be a differentiable function with the property that *supp z* is locally finite (see 2.19), $h > 0$ and the function $\Phi : \boldsymbol{R} \to \boldsymbol{B}_2$ defined in the following way:

$$\Phi(t) = \bigcup_{\xi\in(t-h,t)} z(\xi) \qquad (1)$$

Then $\Phi(t-0)$ exists $\forall t \in \boldsymbol{R}$ and

$$\Phi(t-0) = \bigcup_{\xi\in[t-h,t)} z(\xi) \qquad (2)$$

**Proof** Let us note with $\Omega$ a SINLF family $\omega_s \in \boldsymbol{R}, s \in \boldsymbol{Z}$ satisfying:

$$z(t) = ... \oplus z(\omega_{-1}) \cdot \chi_{\{\omega_{-1}\}}(t) \oplus z(\omega_0) \cdot \chi_{\{\omega_0\}}(t) \oplus z(\omega_1) \cdot \chi_{\{\omega_1\}}(t) \oplus ... \qquad (3)$$

The fact that $supp\ z \subset \Omega$ has the next consequences:

$$\bigcup_{\xi\in(t-h,t)} z(\xi) = \bigcup_{\xi\in(t-h,t)\wedge\Omega} z(\xi) \qquad (4)$$

$$\bigcup_{\xi\in[t-h,t)} z(\xi) = \bigcup_{\xi\in[t-h,t)\wedge\Omega} z(\xi) \qquad (5)$$

Let $t$ arbitrary, but fixed and we remark that for any $\varepsilon > 0$, the sets

$$(t-h-\varepsilon, t-h)\wedge\Omega, [t-\varepsilon, t)\wedge\Omega$$

are finite; we fix $\varepsilon > 0$ sufficiently small in order that they are empty, this fact being always possible.

Let us suppose that $\xi' \in (t-\varepsilon, t)$ is arbitrary. We infer:

$$t-\varepsilon < \xi' < t \Rightarrow t-h-\varepsilon < \xi'-h < t-h \tag{6}$$

$$\Rightarrow (\xi'-h, t-h)\wedge\Omega \subset (t-h-\varepsilon, t-h)\wedge\Omega$$

$$\Rightarrow (\xi'-h, t-h)\wedge\Omega = \emptyset$$

$$\bigcup_{\xi\in(\xi'-h,t-h)\wedge\Omega} z(\xi) = \bigcup_{\xi\in\emptyset} z(\xi) = 0 \tag{7}$$

$$t-\varepsilon < \xi' < t \Rightarrow [\xi', t)\wedge\Omega \subset [t-\varepsilon, t)\wedge\Omega \tag{8}$$

$$\Rightarrow [\xi', t)\wedge\Omega = \emptyset$$

$$\bigcup_{\xi\in[\xi',t)\wedge\Omega} z(\xi) = \bigcup_{\xi\in\emptyset} z(\xi) = 0 \tag{9}$$

and finally, if $h > \varepsilon$, we get $t-h < t-\varepsilon$ and

$$\forall \xi' \in (t-\varepsilon, t), \Phi(\xi') = \bigcup_{\xi\in(\xi'-h,\xi')\wedge\Omega} z(\xi) = \quad \text{(from (4))} \tag{10}$$

$$= \bigcup_{\xi\in(\xi'-h,t-h)\wedge\Omega} z(\xi) \cup \bigcup_{\xi\in[t-h,\xi')\wedge\Omega} z(\xi) = \bigcup_{\xi\in[t-h,\xi')\wedge\Omega} z(\xi) = \quad \text{(from (7))}$$

$$= \bigcup_{\xi\in[t-h,\xi')\wedge\Omega} z(\xi) \cup \bigcup_{\xi\in[\xi',t)\wedge\Omega} z(\xi) = \quad \text{(from (9))}$$

$$= \bigcup_{\xi\in[t-h,t)\wedge\Omega} z(\xi) = \bigcup_{\xi\in[t-h,t)} z(\xi) \quad \text{(from (5))}$$

But the equality between the first and the last member of (10) shows that $\Phi(t-0)$ exists and (2) takes place, i.e. the conclusion of the lemma.

10.4 **Lemma** Let us suppose that $x$ is a solution of Problem 10.1 and for $i \in \{1,...,n\}, t \geq \tau_i$ we have that

$$D^- x_i(t) = 1 \tag{1}$$

Then it is true at least one of the next two possibilities:

a) $\exists j \in \{1,...,n\}, D^- x_j(t-\tau_i) = 1, t-\tau_i \geq \tau_j$ (2)

b) $\exists k \in \boldsymbol{N}, t-\tau_i = \nu_k$ (3)

**Proof** If $t = \tau_i$, then (3) is true under the form

$$t-\tau_i = \nu_0 = 0 \tag{4}$$

thus in (1) we shall suppose that $t > \tau_i$ and we shall prove that

$$D^- f_i(x(t-\tau_i), u(t-\tau_i)) = 1 \tag{5}$$

Case 1, $i \in \{1,...,n_1\}$. (5) is obvious.

Case 2, $i \in \{n_1+1,...,n\}$. We write EAA by taking in consideration (1):

$$x_i(t-0) \oplus x_i(t) = 1 = (x_i(t-0) \oplus f_i(x(t-0), u(t-0))) \cdot \overline{\bigcup_{\xi\in(t-\tau_i,t)} D^- f_i(x(\xi), u(\xi))} \tag{6}$$

resulting that

$$x_i(t-0) \oplus f_i(x(t-0), u(t-0)) = 1 \quad (7)$$

$$\bigcup_{\xi \in (t-\tau_i, t)} D^- f_i(x(\xi), u(\xi)) = 0 \quad (8)$$

We write (6) at the left of $t$ (take a look at 2.21):

$$x_i((t-0)-0) \oplus x_i(t-0) = x_i(t-0) \oplus x_i(t-0) = 0 =$$

$$= (x_i((t-0)-0) \oplus f_i(x((t-0)-0), u((t-0)-0))) \cdot \overline{\bigcup_{\xi \in (t-\tau_i-0, t-0)} D^- f_i(x(\xi), u(\xi))} = \quad (9)$$

$$= (x_i(t-0) \oplus f_i(x(t-0), u(t-0))) \cdot \overline{\bigcup_{\xi \in [t-\tau_i, t)} D^- f_i(x(\xi), u(\xi))}$$

from Lemma 10.3, because the support of $D^- f_i(x(t), u(t))$ is locally finite and if we take in consideration (7),(8), then

$$\bigcup_{\xi \in [t-\tau_i, t)} D^- f_i(x(\xi), u(\xi)) = 1 \quad (10)$$

and (5) is true in this case too.

But $f_i$ is a function and (5) is true if and only if

$$(x(t-\tau_i), u(t-\tau_i)) \neq (x(t-\tau_i-0), u(t-\tau_i-0)) \quad (11)$$

This inequality may be put under the form

$$\exists j \in \{1,...,n\}, x_j(t-\tau_i) \neq x_j(t-\tau_i-0) \quad (12)$$

or

$$\exists j' \in \{1,...,m\}, u_{j'}(t-\tau_i) \neq u_{j'}(t-\tau_i-0) \quad (13)$$

that is equivalent to the conclusion of the lemma that was expressed at a), b) (in (12), the supposition $t-\tau_i > 0$ implies from EAA that $t-\tau_i \geq \tau_j$).

10.5 **Notation** We count the elements of the set

$$\{\nu_k + p_1 \cdot \tau_1 + ... + p_n \cdot \tau_n \mid k, p_1, ..., p_n \in \boldsymbol{N}\}$$

in a strictly increasing order

$$0 = t_0 < t_1 < t_2 < ... \quad (1)$$

and we note with $\boldsymbol{F} = \{t_k \mid k \in \boldsymbol{N}\}$ this SINLF family.

10.6 **Theorem** Let us suppose that Problem 10.1 has solutions and let $x$ be such a solution. Then it is true:

$$supp\, D^- x_1 \vee ... \vee supp\, D^- x_n \subset \boldsymbol{F}$$

**Proof** If the left hand set of the above inclusion is void, then the inclusion is true, so that we shall suppose the contrary: there exist $i_1 \in \{1,...,n\}$ and $t \in \boldsymbol{R}$ so that

$$D^- x_{i_1}(t) = 1 \quad (1)$$

As $t \in (-\infty, 0) \vee (0, \tau_{i_1})$ makes the equality (1) impossible and $t = 0$ belongs to $\boldsymbol{F}$, we shall take in (1) $t \geq \tau_{i_1}$ and this makes Lemma 10.4 possible to be applied under the next form: it is true at least one of

a) $\exists i_2 \in \{1,...,n\}, D^- x_{i_2}(t-\tau_{i_1}) = 1, t-\tau_{i_1} \geq \tau_{i_2}$ (2)

b) $t \in \boldsymbol{F}$

Because b) ends the proof, we shall take in consideration the possibility a). This fact makes possible that Lemma 10.4 be applied again.

In a finite number of steps we reach the situation:

$$t-\tau_{i_1}-\tau_{i_2}-...-\tau_{i_p}=\tau_{i_{p+1}} \quad (3)$$

when $t \in \boldsymbol{F}$ is true too.

10.7 **Corollary** a) Supposing that Problem 10.1 has solutions, any solution $x$ is of the form

$$x(t) = \underset{k \in \boldsymbol{N}}{\Xi}\, x^k \cdot \chi_{[t_k, t_{k+1})}(t) \quad (1)$$

where $\boldsymbol{F} = \{t_k \mid k \in \boldsymbol{N}\}$ is like at 10.5 and $x^k \in \boldsymbol{B}_2^n, k \in \boldsymbol{N}$ are unknown.

b) The next statements are equivalent:

b.1) $x$ is a solution of Problem 10.1

b.2) $x$ satisfies at the time instants $\{t_k \mid k \in \boldsymbol{N}\}$ the equations 10.1 (1), (2) the input $u$ being given by 10.1 (3).

10.8 **Lemma** Let $z \in Real$ of the form

$$z(t) = \underset{k \in \boldsymbol{N}}{\Xi}\, z^k \cdot \chi_{[t_k, t_{k+1})}(t) \quad (1)$$

where $z^k \in \boldsymbol{B}_2, k \in \boldsymbol{N}$ and $t \in \boldsymbol{R}$. Then for any $\tau \in \{\tau_{n_1+1},...,\tau_n\}$ and $t \in \boldsymbol{F}, t \geq \tau$ there are true the formulas:

$$\bigcup_{\xi \in (t-\tau, t)} D^- z(\xi) = \bigcup_{t_s, t_q \in [t-\tau, t) \wedge \boldsymbol{F}} (z^s \oplus z^q) \quad (2)$$

**Proof** Let $t = t_k \in \boldsymbol{F}, t_k \geq \tau$ like in the hypothesis. We have:

$$\bigcup_{\xi \in (t_k-\tau, t_k)} D^- z(\xi) = \begin{cases} 1, \exists \xi \in (t_k-\tau, t_k), z(\xi) \oplus z(\xi-0) = 1 \\ 0, else \end{cases} = \quad (3)$$

$$= \begin{cases} 1, \exists t_s \in (t_k-\tau, t_k) \wedge \boldsymbol{F}, z(t_s) \oplus z(t_{s-1}) = 1 \\ 0, else \end{cases} =$$

$$= \begin{cases} 1, \exists t_s, t_{s-1} \in [t_k-\tau, t_k) \wedge \boldsymbol{F}, z^s \oplus z^{s-1} = 1 \\ 0, else \end{cases} =$$

$$= \begin{cases} 1, \exists t_s, t_q \in [t_k-\tau, t_k) \wedge \boldsymbol{F}, z^s \oplus z^q = 1 \\ 0, else \end{cases} = \bigcup_{t_s, t_q \in [t_k-\tau, t_k) \wedge \boldsymbol{F}} (z^s \oplus z^q)$$

10.9 **Theorem** The Problem 10.1 has a unique solution $x$ of the form 10.7 (1), where $x^k \in \boldsymbol{B}_2^n$ satisfy

$$\forall i \in \{1,...,n_1\}, x_i^{k+1} = \begin{cases} x_i^0, if\ t_{k+1} < \tau_i \\ f_i(x(t_{k+1}-\tau_i), u(t_{k+1}-\tau_i)), if\ t_{k+1} \geq \tau_i \end{cases} \quad (1)$$

$$\forall i\in\{n_1+1,...,n\}, x_i^{k+1}=\begin{cases} x_i^0 & ,if\ t_{k+1}<\tau_i\\ f_i(x(t_k),u(t_k)) & ,if\ t_{k+1}\geq\tau_i\ and\\ \quad \forall t_s,t_q\in[t_{k+1}-\tau_i,t_{k+1})\wedge \boldsymbol{F},\ f_i(x(t_s),u(t_s))=f_i(x(t_q),u(t_q)) & \\ x_i^k & ,if\ t_{k+1}\geq\tau_i\ and\\ \quad \exists t_s,t_q\in[t_{k+1}-\tau_i,t_{k+1})\wedge \boldsymbol{F},\ f_i(x(t_s),u(t_s))\neq f_i(x(t_q),u(t_q)) & \end{cases} \quad (2)$$

for all $k\in \boldsymbol{N}$ .

**Proof** The proof of the theorem means checking the fact that $x$ given by 10.7 (1), where $x^k\in \boldsymbol{B}_2^n, k\in \boldsymbol{N}$ satisfy (1), (2) is a solution of Problem 10.1. Because the first $n_1$ coordinates of $x$ satisfy 10.1 (1), we shall refer to the coordinates $i\in\{n_1+1,...,n\}$ and let us fix such an $i$ arbitrarily.

Case 1, $t\in[0,\tau_i)$. Obvious.

Case 2, $t\geq\tau_i$. We remark, as a consequence of Corollary 10.7 that in this situation it is sufficient to study what happens in an arbitrary $t=t_{k+1}\in \boldsymbol{F}$ . We have from (2):

$$x_i^{k+1}=f_i(x^k,u(t_k))\cdot(1\oplus\bigcup_{t_s,t_q\in[t_{k+1}-\tau_i,t_{k+1})\wedge \boldsymbol{F}}(f_i(x^s,u(t_s))\oplus f_i(x^q,u(t_q))))\oplus \quad (3)$$

$$\oplus x_i^k\cdot\bigcup_{t_s,t_q\in[t_{k+1}-\tau_i,t_{k+1})\wedge \boldsymbol{F}}(f_i(x^s,u(t_s))\oplus f_i(x^q,u(t_q)))=$$

$$=x_i^k\oplus(x_i^k\oplus f_i(x^k,u(t_k)))\cdot(1\oplus\bigcup_{t_s,t_q\in[t_{k+1}-\tau_i,t_{k+1})\wedge \boldsymbol{F}}(f_i(x^s,u(t_s))\oplus f_i(x^q,u(t_q))))$$

Taking in consideration Lemma 10.8 written for

$$z(t)=f_i(x(t),u(t))\cdot\chi_{[0,\infty)}(t), t\in \boldsymbol{R} \quad (4)$$

we see that (3) is equivalent to 10.1 (2), i.e.

$$D^-x_i(t_{k+1})=(x_i(t_{k+1}-0)\oplus f_i(x(t_{k+1}-0),u(t_{k+1}-0)))\cdot$$

$$\cdot(1\oplus\bigcup_{\xi\in(t_{k+1}-\tau_i,t_{k+1})}D^-f_i(x(\xi),u(\xi))) \quad (5)$$

This completes proving that $x$ given by 10.7 (1), with $x^k\in \boldsymbol{B}_2^n, k\in \boldsymbol{N}$ satisfying (1), (2) is a solution of Problem 10.1.

The uniqueness of the solution $x$ has already resulted from the previous reasoning, but we can also prove it directly by supposing against all reason that $y\in Real^{(n)}$ is another solution of EAA. Because

$$\exists t\in \boldsymbol{R}, x(t)\neq y(t) \quad (6)$$

there exists a least $t$ with this property, let it be $t'$. Thus:

$$\forall i\in\{1,...,n\},\forall t\in(-\infty,t'), x_i(t)=y_i(t) \quad (7)$$

$$\exists j\in\{1,...,n\}, x_j(t')\neq y_j(t') \quad (8)$$

If $j\in\{1,...,n_1\}$, the contradiction is obvious and we shall suppose now that $j\in\{n_1+1,...,n\}$. We can write, taking in consideration that (8) implies $t'\geq\tau_j$:

$$x_j(t')=x_j(t'-0)\oplus(x_j(t'-0)\oplus f_j(x(t'-0),u(t'-0)))\cdot\overline{\bigcup_{\xi\in(t'-\tau_j,t')}D^- f_j(x(\xi),u(\xi))}= \quad (9)$$

$$=y_j(t'-0)\oplus(y_j(t'-0)\oplus f_j(y(t'-0),u(t'-0)))\cdot\overline{\bigcup_{\xi\in(t'-\tau_j,t')}D^- f_j(y(\xi),u(\xi))}=y_j(t')$$

contradiction with (8), showing that $x$ and $y$ are equal.

10.10 **Problem** To be solved the system of equations EAAA

$$x_i(t)=g_i(x(t-\tau_i))\cdot\chi_{[\tau_i,\infty)}(t)\oplus x_i^0\cdot\chi_{[0,\tau_i)}(t),i=\overline{1,n_1} \quad (1)$$

$$D^- x_i(t)=(x_i(t-0)\oplus g_i(x(t-0)))\cdot \quad (2)$$

$$\cdot\overline{\bigcup_{\xi\in(t-\tau_i,t)}D^- g_i(x(\xi))}\cdot\chi_{[\tau_i,\infty)}(t)\oplus x_i^0\cdot\chi_{\{0\}}(t),i=\overline{n_1+1,n}$$

with $g:\mathbf{B}_2^n\to\mathbf{B}_2^n$ and $t\in\mathbf{R}$; $x^0\in\mathbf{B}_2^n$ and $\tau_1,...,\tau_n>0$ are given and $x\in Real^{(n)}$ is the unknown.

It is supposed, without loss, that $1\le n_1<n$.

10.11 **Remark** Solving the Problem 10.10 can be considered to be a special case of solving the Problem 10.1, that is the case when the asynchronous automaton is autonomous-like and we have

$$u^0=u^1=u^2=... \quad (1)$$

the SINLF family $(\nu_k)$ being arbitrary. The relation between $f$ and $g$ is given by:

$$f(\cdot,u^0)=g(\cdot) \quad (2)$$

10.12 **Notation** We count the elements of the set $\{p_1\cdot\tau_1+...+p_n\cdot\tau_n\mid p_1,...,p_n\in\mathbf{N}\}$ in a strictly increasing order:

$$0=t_0^{'}<t_1^{'}<t_2^{'}<... \quad (1)$$

and we note with $\mathbf{F}^0=\{t_k^{'}\mid k\in\mathbf{N}\}$ this SINLF family.

10.13 **Remark** Of course that the set $\mathbf{F}^0$ results from the wish of making $\mathbf{F}$ as small as possible by a suitable choice of $(\nu_k)$, which is arbitrary. We have chosen

$(\nu_k)\subset\{p_1\cdot\tau_1+...+p_n\cdot\tau_n\mid p_1,...,p_n\in\mathbf{N}\}$.

10.14 **Corollary** of Theorem 10.9. The Problem 10.10 has a unique solution $x$ that may be put under the form

$$x(t)=\underset{k\in\mathbf{N}}{\Xi}\, x^k\cdot\chi_{[t_k^{'},t_{k+1}^{'})}(t) \quad (1)$$

where $x^k\in\mathbf{B}_2^n$ satisfy

$$\forall i\in\{1,...,n_1\},x_i^{k+1}=\begin{cases}x_i^0 & ,if\ t_{k+1}^{'}<\tau_i\\ g_i(x(t_{k+1}^{'}-\tau_i)),if\ t_{k+1}^{'}\ge\tau_i\end{cases} \quad (2)$$

$$\forall i \in \{n_1+1,...,n\}, x_i^{k+1} = \begin{cases} x_i^0 & ,if\ t'_{k+1} < \tau_i \\ g_i(x(t'_k)) & ,if\ t'_{k+1} \geq \tau_i\ and \\ \quad \forall t'_s, t'_q \in [t'_{k+1} - \tau_i, t'_{k+1}) \wedge \boldsymbol{F}^0, g_i(x(t'_s)) = g_i(x(t'_q)) & \\ x_i^k & ,if\ t'_{k+1} \geq \tau_i\ and \\ \quad \exists t'_s, t'_q \in [t'_{k+1} - \tau_i, t'_{k+1}) \wedge \boldsymbol{F}^0, g_i(x(t'_s)) \neq g_i(x(t'_q)) & \end{cases} \quad (3)$$

for all $k \in \boldsymbol{N}$.

10.15 **Remark** There are three time periods that characterize the trajectory of the automaton $\Sigma$:

- $t < 0$, when the automaton *did not start yet*
- $t \in [0, \tau_i)$, when the automaton *gets initialized* on the $i-th$ coordinate
- $t \geq \tau_i$, the deterministic run, $i \in \{1,...,n\}$.

10.16 **Corollary** of Theorem 10.9. The trajectory $x$ of $\Sigma$ satisfies the next properties:

$$D^- x_i(t) = 1 \Rightarrow D^- f_i(x(t-\tau_i), u(t-\tau_i)) = 1, t > \tau_i, i = \overline{1,n} \quad (1)$$

$$\bigcup_{\xi \in (t-\tau_i, t)} D^- f_i(x(\xi), u(\xi)) = 0 \Rightarrow x_i(t) = f_i(x(t-\tau_i), u(t-\tau_i)), t \geq \tau_i, i = \overline{n_1+1, n} \quad (2)$$

$$(t < t'\ and\ D^- x_i(t) = D^- x_i(t') = 1) \Rightarrow t' - t \geq \tau_i, i = \overline{n_1+1, n} \quad (3)$$

**Proof** The idea of showing the validity of (1) has appeared in the proof of 10.4 and (2) is easily proved.

We prove (3). If $t = 0$, then it is clear that $t' \geq \tau_i$ and the conclusion of (3) is true. We take $t, t' \geq \tau_i$. From EAA it results that:

$$\bigcup_{\xi \in (t-\tau_i, t)} D^- f_i(x(\xi), u(\xi)) = 0 \quad (4)$$

Let us suppose against all reason that we have

$$t' - t < \tau_i \quad (5)$$

Because there are true

$$t - \tau_i < t' - \tau_i < t \quad (6)$$

$$D^- f_i(x(t'-\tau_i), u(t'-\tau_i)) = 1 \quad \text{(from (1))} \quad (7)$$

we obtain that

$$\bigcup_{\xi \in (t-\tau_i, t)} D^- f_i(x(\xi), u(\xi)) \geq D^- f_i(x(t'-\tau_i), u(t'-\tau_i)) = 1 \quad (8)$$

and this in contradiction with (4). We have that (5) is false.

10.17 **Remark** At 10.16, (1) and (2) are properties of determinism, interpreted like this:

- the effect $D^- x_i(t) = 1$ has been caused by $D^- f_i(x(t-\tau_i), u(t-\tau_i)) = 1$
- if the cause $f_i(x(\cdot), u(\cdot))$ is constant long enough at $t$, then it has the effect $x_i(t) = f_i(x(t-\tau_i), u(t-\tau_i))$

and (3) is a property of inertiality: $x_i$ cannot switch sooner than once at each $\tau_i$ time units. The first statements of this nature were made at 5.5.

**11. Continuous Time and Discrete Time**

11.1 **Remark** The coexistence of the real time with the discrete time has appeared from the beginning of this work. Thus, the differentiable functions that have been defined in paragraph 2 are real time functions but, as we can see from Theorem 2.16, we can choose for any $x \in Diff$ a new time set $\{t_z \mid z \in \boldsymbol{Z}\}$.

11.2 **Definition** Let $x \in Diff$ . We call *essential time set* of $x$ the set $T_x$ that is defined by:

$$T_x = supp\ D^- x \vee supp\ D^+ x \tag{1}$$

More general, the *essential time set* of $x \in Diff^{(n)}$ is defined by:

$$T_x = T_{x_1} \vee ... \vee T_{x_n} \tag{2}$$

11.3 **Remark** If $x \in Diff$ is constant, $T_x = \emptyset$; if $x$ is *monotonous*, i.e. if it has on the real axis exactly one switch from 0 to 1 (increasingly monotonous) or one switch from 1 to 0 (decreasingly monotonous) in $t_0$, then $T_x = \{t_0\}$. Otherwise, $T_x$ can be finite, it can be of the form $T_x = \{t_z \mid z \in \boldsymbol{N}\}$ for some realizable (or realizable*) functions, or of the form $T_x = \{t_z \mid z \in \boldsymbol{Z}\}$ for some differentiable functions. Each time, we can substitute the time set $\boldsymbol{R}$ with the time set $T_x$ or some other locally finite (see 2.16 (2)) set $T \supset T_x$. $T$ itself can be identified with any ordered finite set, with $\boldsymbol{N}$ or with $\boldsymbol{Z}$ .

We have started the analysis of EAA from the time set $\boldsymbol{R}$ and we have reached the time set $\boldsymbol{F}$ . Similarly, the analysis of EAAA starts from $\boldsymbol{R}$ and gets to $\boldsymbol{F}^0$ .

11.4 **Remark** The fact that the delays of the asynchronous automata are not known (except for the fixed delay model) makes that the "sampling moments" $\{t_k \mid k \in \boldsymbol{N}\}$ are not known themselves and this gives another perspective on the discrete time.

11.5 **Remark** The relation real time-discrete time is a rather complex one and some of the remarkable mathematicians that have papers in this topic are M. Vardi, T. Henzinger, A. Pnuelli, R. Alur. We quote from the introduction of [Luca, Manna, 1995]:

"There are two common choices for the semantics of real-time systems. The first is a discrete semantics, in which the temporal evolution of the system is represented as an enumerable sequence of snapshots, each describing the state of the system at a certain time. The second is a continuous semantics, in which the system evolution is represented by a sequence of intervals of time, together with a description of the system state during each interval".

In order to relate the real time to the discrete time, the authors of the cited paper state the hypothesis of finite variability FV of the formulas (that we do not define here) that is quite similar to our condition of local finiteness. We just mention that FV is thought by Alfaro and Manna so that the validity of a formula in the discrete semantics implies its validity in the continuous one. Our condition of local finiteness has its origin in our desire of modeling the electrical signals.

**12. Transitions. The Interleaving Concurrency Model**

12.1 **Remark** This paragraph is dedicated to the discrete time systems and the previous statements concerning the relation discrete time - continuous time are kept in mind. Some more insight in branching time will be made in the paragraph 19.

12.2 **Notation** We note with $\varepsilon^i = (0,0,...,\underset{i}{1},...,0) \in \mathbf{B}_2^{m+n}, i = \overline{1, m+n}$ the vectors of the canonical base of the linear space $\mathbf{B}_2^{m+n}$.

12.3 **Remark** If $w, w' \in \mathbf{B}_2^{m+n}$ are two arbitrary vectors, the meaning of the sum:

$$w \oplus w' = \varepsilon^{i_1} \oplus ... \oplus \varepsilon^{i_p}, i_1,...,i_p \in \{1,...,m+n\} \qquad (1)$$

is that $w$ and $w'$ differ on the coordinates $i_1,...,i_p$.

12.4 **Definition** Let $A \subset \mathbf{B}_2^{m+n} \times \mathbf{B}_2^{m+n}$ a relation. The *domain* of $A$ is defined by:

$$dom\ A = \{w \mid w \in \mathbf{B}_2^{m+n}, \exists w' \in \mathbf{B}_2^{m+n}, (w, w') \in A\} \qquad (1)$$

12.5 **Definition** Let $u^0, u^1, ... \in \mathbf{B}_2^m$, $x^0 \in \mathbf{B}_2^n$, $f : \mathbf{B}_2^m \times \mathbf{B}_2^n \to \mathbf{B}_2^n$ and $m \geq 0, n \geq 1$. The set of the *transitions* (or of the *transfers*) $\Gamma \subset \mathbf{B}_2^{m+n} \times \mathbf{B}_2^{m+n}$ is defined like this:

a) $(u^0, x^0) \in dom\ \Gamma$

b) we suppose that $(u^k, x) \in dom\ \Gamma$. If

$$(u^k, x) \oplus (u^{k+1}, f(x, u^k)) = a_1^k \cdot \varepsilon^1 \oplus ... \oplus a_{m+n}^k \cdot \varepsilon^{m+n} \qquad (1)$$

and $a_1,...,a_{m+n} \in \mathbf{B}_2$ satisfy $a_1 \leq a_1^k,...,a_{m+n} \leq a_{m+n}^k$, then

b.1) $(u^k, x) \oplus a_1 \cdot \varepsilon^1 \oplus ... \oplus a_{m+n} \cdot \varepsilon^{m+n} \in dom\ \Gamma$

b.2) $((u^k, x), (u^k, x) \oplus a_1 \cdot \varepsilon^1 \oplus ... \oplus a_{m+n} \cdot \varepsilon^{m+n}) \in \Gamma$

b.3) $(y, y'), (y', y'') \in \Gamma \Rightarrow (y, y'') \in \Gamma$

c) all the elements of $\Gamma$ are given by a), b).

12.6 **Remark** We interpret Definition 12.5 in a systemic manner, $u^0, u^1, ... \in \mathbf{B}_2^m$ being the sequence of the input values and $x^0 \in \mathbf{B}_2^n$ respectively the initial state of an asynchronous automaton $\Sigma$, like in paragraph 10 for example.

12.7 **Definition** The transition $(y, y) \in \Gamma$ is called *trivial*, the transition $(y, y \oplus \varepsilon^i) \in \Gamma$ is called *elementary* and the transition $(y, y'') \in \Gamma$ from 12.5 b.3) is said to result by the *composition* of the transitions $(y, y')$ and $(y', y'')$, in this order.

12.8 **Definition** The elements $y \in dom\ \Gamma$, as well as the couples (input, output) $(u, x) \in Real^{(m+n)}$ are called the *extended state*, or the *total state* of $\Sigma$.

12.9 **Notation** An alternative notation for $(y, y') \in \Gamma$ is $y \to y'$.

12.10 **Definition** In the equations 12.5 (1), the coordinates $i \in \{1,...,m+n\}$ for which

$$a_i^k = 1 \tag{1}$$

are called *excited* and the rest of the coordinates are called *stable*, $k = 0,1,2,...$

12.11 **Notation** It is usual to note with an asterisk '*' the excited coordinates.

12.12 **Definition** We use to say about the non-trivial transitions $y \to y'$ that they are *enabled* (meaning that they can happen) and about a certain non-trivial transition that just happens - could it be elementary or not - that it *fires*.

12.13 **Definition** The hypothesis of *interleaving concurrency* states that any time a transition fires, it is elementary.

12.14 **Example** In the next drawing

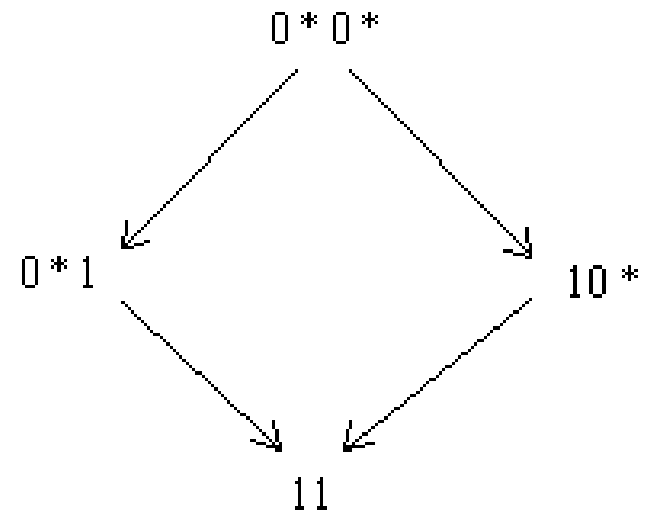

fig (1)

(called *state transition diagram*), the first coordinate is the input and the second is the state:

$$u(t) = \chi_{[\nu,\infty)}(t) \tag{2}$$

$$x(t) = \chi_{[t_1,\infty)}(t) \tag{3}$$

with $\nu, t_1 \geq 0$. We have:

$$f(0,0) = f(0,1) = f(1,0) = f(1,1) = 1 \tag{4}$$

and the hypothesis of interleaving concurrency states that $\nu \neq t_1$, resulting:

$$y^0 = (0,0) \tag{5}$$

$$y^2 = (1,1) \tag{6}$$

If $\nu < t_1$, then

$$y^1 = (1,0) \tag{7}$$

and if $\nu > t_1$, then

$$y^1 = (0,1) \tag{8}$$

$\Gamma$ has five elements:

$$\Gamma = \{((0,0),(0,1)),((0,0),(1,0)),((0,0),(1,1)),((0,1),(1,1)),((1,0),(1,1))\} \tag{9}$$

At the initial time moment three transitions are enabled and only two can fire.

12.15 **Definition** We call a *race* the situation when some transition $(y, y \oplus \varepsilon^i \oplus ... \oplus \varepsilon^j) \in \Gamma$ fires without obeying the interleaving concurrency model, i.e. if several coordinates $i,...,j \in \{1,...,n\}$ switch at the same time. We say that the switching coordinates *have won the race* (with the other enabled coordinates, that did not switch).

12.16 **Remark** The excited coordinates show the direction towards which a system runs; the race winner depends on the values of $\tau_1,...,\tau_n$.

12.17 **Remark** The hypothesis of interleaving concurrency (giving the so called *interleaving concurrency model*) is based on the fact that in nature, perfect simultaneity does not exist. It is correct in some sense, but it refers to a special case that we shall not adopt with the exception of the situations when we shall explicitly mention the contrary.

12.18 **Remark** Our "extended state" may be associated to the "state" of Lavagno, when the condition of injectivity of the labeling function (from the set of the states to the "modeling space" $\boldsymbol{B}_2^n$) is fulfilled. What we call "coordinate of the extended state $y_i$", he calls "signal" and his demand is that "a single signal changes for every transition". For him, interleaving concurrency means "considering all possible alternative chain orderings compliant with the partial order between possibly concurrent actions".

For Kondratyev et al. "a transition between states is a transition of exactly one signal. There may be many signals enabled in a state, but exactly one signal transition is fired at a time. This corresponds to the interleaving concurrency model". What they mean by "state" and "signal" is close enough to what we mean by "extended state" and "coordinate of the extended state".

## 13. Points of Equilibrium. The Stability

13.1 **Notations** We continue to refer to the generator function $g$ from the definition 6.3; we note with $x$, see 10.14 (1), the solution of the Problem 10.10 - the fixed delay model was used there - and with $\Sigma$ the appropriate autonomous automaton.

$\overline{\varepsilon}^i = (0,0,...,\underset{i}{1},...,0) \in \boldsymbol{B}_2^n, i = \overline{1,n}$ are the vectors of the canonical base.

13.2 **Remark** The next definition adapts 12.10 to the present requests: passing

- from the discrete time to the continuous time
- from the non-autonomous automata to the autonomous automata

13.3 **Definition** Let $t \ge 0$ given and $k \in \boldsymbol{N}$ so that $t \in [t_k, t_{k+1})$. In the equations

$$x^k \oplus g(x^k) = a_1^k \cdot \overline{\varepsilon}^1 \oplus ... \oplus a_n^k \cdot \overline{\varepsilon}^n \tag{1}$$

the coordinates $i \in \{1,...,n\}$ for which

$$a_i^k = 1 \tag{2}$$

are called *excited* at $t$ and the rest of the coordinates are called *stable* at $t$.

13.4 **Theorem** Let us consider EAAA. The next statements are equivalent:

a) $$g(x^0) = x^0 \tag{1}$$

b) $x_i, i = \overline{1,n}$ are stable at 0

c) $$x(t) = x^0 \cdot \chi_{[0,\infty)}(t) \tag{2}$$

d) $$g_i(x^0) = x_i^0, i = \overline{1,n_1} \tag{3}$$

$$D^- x_i(\tau_i) = 0, i = \overline{n_1 + 1, n} \tag{4}$$

**Proof** a) $\Leftrightarrow$ b) is obvious. a) $\Rightarrow$ c) results from 10.14: at all time instants $t_0^{'}, t_1^{'}, t_2^{'}, ...$ see 10.12 and for all $i = \overline{1,n}$, we have

$$x_i^{k+1} = x_i^0 \tag{5}$$

c) $\Rightarrow$ d) By replacing the form of $x$ from (2) in 10.14 (1), we have that 10.14 (2) implies the validity of (3); (2) $\Rightarrow$ (4) is obvious.

d) $\Rightarrow$ a) $\tau_{n_1+1},...,\tau_n$ written without repetitions, in a strictly increasing order, are noted with $t_{k_1}^{'} < ... < t_{k_p}^{'}$. The equations

$$\forall i \in \{j \mid j \in \{n_1+1,...,n\}, \tau_j = t_{k_1}^{'}\}, \begin{cases} x_i(\tau_i) = x_i^0 & \text{, from (4)} \\ x_i(\tau_i) = g_i(x^0), & \text{from 10.14 (3)} \end{cases} \tag{6}$$

$$...$$

$$\forall i \in \{j \mid j \in \{n_1+1,...,n\}, \tau_j = t_{k_p}^{'}\}, \begin{cases} x_i(\tau_i) = x_i^0 & \text{, from (4)} \\ x_i(\tau_i) = g_i(x^0), & \text{from (6), ... and from 10.14 (3)} \end{cases} \tag{7}$$

give the validity of a).

13.5 **Remark** One of the interpretations that we can give to the Theorem 13.4 is the following one: if $x_i, i = \overline{1,n}$ are stable at 0, then they are stable at any $t \geq 0$.

13.6 **Definition** If one of the conditions 13.4 a),...,d) is satisfied, we say that the autonomous automaton $\Sigma$ is *trivially stable*.

13.7 **Remark** Let us point out the situation when in 13.4

$$x^0 = 0 \tag{1}$$

This is the trivial stability of the trivial autonomous automata.

13.8 **Theorem** In EAAA, let $x' \in \boldsymbol{B}_2^n, t' \geq \max\{\tau_1,...,\tau_n\}$ and we suppose that

$$g_i(x(t)) = x_i^{'}, t \in [t'-\tau_i, t'), i = \overline{1,n} \tag{1}$$

The next statements are equivalent:

a) $$g(x') = x' \tag{2}$$

b) $x_i, i = \overline{1,n}$ are stable at $t'$

c) $$x(t) = x', t \geq t' \tag{3}$$

d) $$g_i(x') = x_i^{'}, i = \overline{1,n_1} \tag{4}$$

$$D^- x_i(t') = 0, i = \overline{n_1+1,n} \tag{5}$$

**Proof** We can suppose without loss that $t' \in \boldsymbol{F}^0$ (if not, we can replace $t'$ with the smallest $t'' \in \boldsymbol{F}^0$ so that $t' < t''$).

a) $\Rightarrow$ b) 10.14 gives

$$x(t') = x' = g(x') = g(x(t')) \tag{6}$$

The implications b) $\Rightarrow$ c), c) $\Rightarrow$ d), d) $\Rightarrow$ a) are easily proved.

13.9 **Remark** Similarly to 13.5, the Theorem 13.8 may be interpreted like this. If 13.8 (1) is true (this hypothesis is analogue to the initializing of $x_i$ on the $[0,\tau_i)$ interval from 13.4, that is understood there and explicit here), then the stability of $x_i, i=\overline{1,n}$ at $t'$ implies the stability at any $t \geq t'$.

13.10 **Definition** If $x', t'$ exist so that the autonomous automaton $\Sigma$ satisfies 13.8 (1) together with one of the conditions a),...,d), then we say that it is *stable*. If $\Sigma$ is not stable, it is called *unstable*.

13.11 **Remark** If $\Sigma$ is trivially stable, it is stable. If it is not trivially stable, then it is either stable with $t' > 0$, or unstable.

13.12 **Remark** We use to identify the stability of the autonomous automata with the existence of $t'$ and of the fix point $x' \in \boldsymbol{B}_2^n$ of $g$, so that

$$\forall t \geq t', x(t) = x' \tag{1}$$

Then the instability can be interpreted like this:

$$\forall t', \exists t > t', x(t) \neq x(t') \tag{2}$$

13.13 **Definition** The vector $x'$ is called *point of equilibrium*, or *stable state*, or *steady state* (*of the trajectory* $x$, or *of the automaton* $\Sigma$).

A value that $x$ takes which is not a point of equilibrium is called *unstable*, or *transient*.

13.14 **Remark** The notion of point of equilibrium belongs to the field theory, see for example [Udriste, 1988] where the differential equations are written on real numbers. In this context, $g$ is called (*Boolean*) *vector field* and $x$ is called (*pseudoboolean*) *field line* (of $g$).

The terminology of stable state appears is asynchronous automata theory and steady states appear in the more general frame given by the systems theory.

13.15 **Notations** We shall refer from this moment to the non-autonomous asynchronous automaton $\Sigma$, with the input $u$ given by 10.1 (3); $f$ is its generator function, $x$ is the solution of the Problem 10.1 and $y$ is the extended state.

13.16 **Definition** (see 13.3) Let $t \geq 0$ given and $k \in \boldsymbol{N}$ with the property that $t \in [t_k, t_{k+1})$. In the equations

$$x^k \oplus f(x^k, u(t_k)) = a_1^k \cdot \overline{\varepsilon}^1 \oplus ... \oplus a_n^k \cdot \overline{\varepsilon}^n \tag{1}$$

the coordinates $i \in \{1,...,n\}$ for which

$$a_i^k = 1 \tag{2}$$

are called *excited at t*. The coordinates that are not excited at $t$ are called *stable at t*.

13.17 **Theorem** Let EAA - the fixed delay model - and we suppose that $\nu_{k+1} - \nu_k \geq \max\{\tau_1,...,\tau_n\}, k \in \boldsymbol{N}$. The next statements are equivalent:

a) $$\forall k \in \boldsymbol{N}, f(x^0, u^k) = x^0 \tag{1}$$

b) $x_i, i=\overline{1,n}$ are stable at any $\nu_k, k \in \boldsymbol{N}$

c) $$x(t) = x^0 \cdot \chi_{[0,\infty)}(t) \tag{2}$$

d) $$\forall k\in \boldsymbol{N},\begin{cases} f_i(x^0,u^k)=x_i^0 & ,i=\overline{1,n_1} \\ D^- x_i(\nu_k+\tau_i)=0, i=\overline{n_1+1,n} \end{cases} \quad (3)$$

13.18 **Definition** If one of the conditions 13.17 a),...,d) is true, then we say that the non-autonomous automaton $\Sigma$ is *trivially stable*.

13.19 **Remark** The special case when

$$x^0=0 \quad (1)$$

gives in 13.17 the trivial stability of the non-autonomous trivial automata.

13.20 **Theorem** We suppose the existence of $x'\in \boldsymbol{B}_2^n$ and of $t'\geq \max\{\tau_1,...,\tau_n\}$ so that in EAA we have $\nu_{k+1}-\nu_k\geq \max\{\tau_1,...,\tau_n\}, k\in \boldsymbol{N}$ and:

$$f_i(x(t),u(t))=x_i^{'}, t\in [t'-\tau_i,t'), i=\overline{1,n} \quad (1)$$

The next statements are equivalent:

a) $$\forall k,\nu_k\geq t'\Rightarrow f(x',u^k)=x' \quad (2)$$

b) $x_i, i=\overline{1,n}$ are stable at any $\nu_k\geq t'$

c) $$x(t)=x', t\geq t' \quad (3)$$

d) $$\forall k,\nu_k\geq t'\Rightarrow \begin{cases} f_i(x',u^k)=x_i^{'} & ,i=\overline{1,n_1} \\ D^- x_i(\nu_k+\tau_i)=0, i=\overline{n_1+1,n} \end{cases} \quad (4)$$

13.21 **Definition** If the non-autonomous automaton $\Sigma$ satisfies 13.20 (1) and one of the conditions 13.20 a),...,d) we say that it is *stable*. If $\Sigma$ is not stable, then it is called *unstable*.

13.22 **Remark** We identify the stability of the non-autonomous automaton with the existence of $t'\geq 0$ and $x'\in \boldsymbol{B}_2^n$ with the property that $x'$ is a fix point for all the functions $f(\cdot,u^k)$, where $\nu_k\geq t'$ and we have

$$\forall t\geq t', x(t)=x' \quad (1)$$

The meaning of the instability is given by

$$\forall t',\exists t>t', x(t)\neq x(t') \quad (2)$$

13.23 **Remark** The vector $x'$ may be called once more point of equilibrium, even if this definition is more natural for the autonomous automata and for the autonomous-like automata, when

$$u(t)=u^0\cdot \chi_{[0,\infty)}(t) \quad (1)$$

and $f$ is called *vector field with parameter* (the parameter is $u^0$).

13.24 **Remark** The previous requests related to the stability of the non-autonomous automata refer to $x$; if we replace the state $x$ with the extended state $y$, it will result a stronger concept of stability, i.e.: we add to 13.22 (1) the request

$$\forall t\geq t', u(t)=u' \quad (1)$$

It will always result to which definition we refer.

13.25 **Examples** The clock generator, the automaton from 7.1 (1) is unstable, 13.12 (2) being satisfied and this has generated the situation when modeling was not possible using the unbounded delay model.

The R-S latch, the autonomous-like automaton from 8.1 (1) is stable in all the situations with the exceptions

iv) j) for the pure-delay model

iv) b) j) for the inertial delay model

For this automaton, the cases i), ii), iii) imply the stability and 8.3 (2), (3) show that modeling is possible using the unbounded delay model.

## 14. The Fundamental Mode of Operation

14.1 **Informal definition** A non-autonomous asynchronous automaton $\Sigma$ is considered. We say that it operates in the fundamental mode, if

a) [Lavagno, 1992] "inputs are considered to change only when all the delay elements are stable (i.e. they have the input value equal to the output value)"

b) [Kishinevsky+,1997] we have a "slow enough environment" "(inputs can change **after** the system has settled into a stable state)".

14.2 **Remark** We make the following remarks on the previous informal definitions.

By stability identified with the situation when "input value is equal to the output value" is understood the fact that at $t'$, the trajectory $x$ has reached a value $x'$ that is a fix point of $f(\cdot,u(t'))$: as argument of this function $x'$ is "input value" and as value of this function $x'$ is "output value".

The environment is associated to the input and the authors identify them. This point of view is related rather to modeling a circuit than to controlling a process, when the input may be called control.

14.3 **Definition** Let us consider the fixed delay model and we suppose that the asynchronous non-autonomous automaton $\Sigma$ satisfies one of the next two conditions:

a) $$u(t)=u^0\cdot\chi_{[\nu_0,\nu_1)}(t)\oplus u^1\cdot\chi_{[\nu_1,\nu_2)}(t)\oplus...\oplus u^k\cdot\chi_{[\nu_k,\infty)}(t) \quad (1)$$

where

$$0=\nu_0<\nu_1<...<\nu_k \quad (2)$$

and the numbers $t_0,t_1,...,t_k$ exist so that

$$\nu_0<t_0\le\nu_1<t_1\le\nu_2<...\le\nu_k<t_k \quad (3)$$

and $x_i, i=\overline{1,n}$ are stable at $t_0,t_1,...,t_k$

b) $$u(t)=u^0\cdot\chi_{[\nu_0,\nu_1)}(t)\oplus u^1\cdot\chi_{[\nu_1,\nu_2)}(t)\oplus...\oplus u^k\cdot\chi_{[\nu_k,\nu_{k+1})}(t)\oplus... \quad (4)$$

where $\nu_k, k\in\mathbf{N}$ is SINLF and the family $t_k, k\in\mathbf{N}$ exists so that

$$\nu_0<t_0\le\nu_1<t_1\le\nu_2<...\le\nu_k<t_k\le... \quad (5)$$

and $x_i, i=\overline{1,n}$ are stable at $t_0,t_1,...,t_k,...$

Then we say that $\Sigma$ operates in the *fundamental mode*.

14.4 **Remark** A situation of triviality is possible for the fundamental mode, when $\exists k$ so that $x_i, i=\overline{1,n}$ are stable at $\nu_k$. This is the case for example if $u^{k-1}=u^k$.

14.5 **Remark** A condition that we have not put at 14.3 is of the type $\nu_{k+1} - \nu_k \geq \max\{\tau_1,...,\tau_n\}$, see also 13.17, 13.20.

14.6 **Remark** Any stable autonomous-like automaton operates in the fundamental mode (the definition 14.3 a) for $k = 0$). Such an example has been given at 8.1, the R-S latch, in the situations when it is stable.

14.7 **Remark** The fundamental mode represents a relation between the environment and the system, more precisely between the speed of variation of the input and the speed of stabilization of the system.

We are tempted to make an analogy here. The input is the professor that sends information and the system is the group of students that receives the information. The students understand what the professor teaches - $x_i, i = \overline{1,n}$ are stable at $t_0, t_1,..., t_k$ - if the next two conditions are fulfilled:

- the information is not contradictory, i.e. the stability of $x_i, i = \overline{1,n}$ is possible
- the information is sent slowly enough so that the students have time to understand (to stabilize).

14.8 **Remark** The definition 14.3 must be understood to be dependent on the delay model that we use; for the unbounded delay model 5.8 c"), the definition item a) becomes: "the numbers $t_0, t_1,..., t_k$ exist so that for any $\tau_1 \in (0, M_1],..., \tau_n \in (0, M_n]$ 14.3 (3) is true and $x_i, i = \overline{1,n}$ are stable at $t_0, t_1,..., t_k$".

The way that this definition must be given in the other cases is obvious now.

## 15. Combinational Automata

15.1 **Notations** Let $f : \mathbf{B}_2^n \times \mathbf{B}_2^m \to \mathbf{B}_2^n$, $g : \mathbf{B}_2^n \to \mathbf{B}_2^n$ the generator functions of an asynchronous automaton $\Sigma$, in the non-autonomous and the autonomous version. If $\Sigma$ is combinational, i.e. if the next properties are true:

$$\forall x, \forall x', f(x, \cdot) = f(x', \cdot) \tag{1}$$

$$\forall x, \forall x', g(x) = g(x') \tag{2}$$

then the two functions are noted with $f : \mathbf{B}_2^m \to \mathbf{B}_2^n$ and $c \in \mathbf{B}_2^n$ (the constant function, identified with the constant).

15.2 **Remark** The equations of the combinational automata are of the form:

$$x_i(t) = f_i(u(t - \tau_i)) \cdot \chi_{[\tau_i,\infty)}(t) \oplus x_i^0 \cdot \chi_{[0,\tau_i)}(t), i = \overline{1, n_1} \tag{1}$$

$$D^- x_i(t) = (x_i(t-0) \oplus f_i(u(t-0))) \cdot \tag{2}$$

$$\cdot \overline{\bigcup_{\xi \in (t-\tau_i, t)} D^- f_i(u(\xi))} \cdot \chi_{[\tau_i,\infty)}(t) \oplus x_i^0 \cdot \chi_{\{0\}}(t), i = \overline{n_1 + 1, n}$$

for the non-autonomous case, respectively of the form

$$x_i(t) = c_i \cdot \chi_{[\tau_i,\infty)}(t) \oplus x_i^0 \cdot \chi_{[0,\tau_i)}(t), i = \overline{1, n_1} \tag{3}$$

$$D^- x_i(t) = (x_i(t-0) \oplus c_i) \cdot \chi_{[\tau_i,\infty)}(t) \oplus x_i^0 \cdot \chi_{\{0\}}(t), i = \overline{n_1 + 1, n} \tag{4}$$

for the autonomous case. Let us observe the formal coincidence between (3) and (4) meaning that the autonomous combinational automata do not have the set of the state coordinates partitioned in pure and inertial coordinates.

15.3 **Theorem** Let the combinational automaton $\Sigma$ be associated to the combinational circuit $\tilde{\Sigma}$ (see 9.5). We suppose that its input $u$ is given by 14.3 (1) and let the equations:

$$x^0 \oplus f(u^0) = a_1^0 \cdot \bar{\varepsilon}^{-1} \oplus ... \oplus a_n^0 \cdot \bar{\varepsilon}^{-n} \quad (1)$$

$$f(u^0) \oplus f(u^1) = a_1^1 \cdot \bar{\varepsilon}^{-1} \oplus ... \oplus a_n^1 \cdot \bar{\varepsilon}^{-n} \quad (2)$$

$$...$$

$$f(u^{k-1}) \oplus f(u^k) = a_1^k \cdot \bar{\varepsilon}^{-1} \oplus ... \oplus a_n^k \cdot \bar{\varepsilon}^{-n} \quad (3)$$

where $a_1^j,...,a_n^j \in \mathbf{B}_2, j = \overline{0,k}$. We note

$$\varphi_j = \max\{\tau_i \mid i \in \{1,...,n\}, a_i^j = 1\}, \ j = \overline{0,k} \quad (4)$$

$$\Theta_j = \max\{M_i \mid i \in \{1,...,n\}, a_i^j = 1\}, \ j = \overline{0,k} \quad (5)$$

$M_i$ being the superior estimates of $\tau_i, i = \overline{1,n}$ in the unbounded delay model 5.8 c").

a) If, in the fixed delay model, we have

$$\nu_1 - \nu_0 \geq \varphi_0 \quad (6)$$

$$\nu_2 - \nu_1 \geq \varphi_1 \quad (7)$$

$$...$$

$$\nu_k - \nu_{k-1} \geq \varphi_{k-1} \quad (8)$$

then the solution $x$ of 15.2 (1), (2) satisfies that $x_i, i = \overline{1,n}$ are stable at $\nu_0 + \varphi_0, \nu_1 + \varphi_1,...$ $\nu_k + \varphi_k$, $\Sigma$ operates in the fundamental mode and moreover

$$\forall t \in [\nu_0 + \varphi_0, \nu_1], x(t) = f(u^0) \quad (9)$$

$$\forall t \in [\nu_1 + \varphi_1, \nu_2], x(t) = f(u^1) \quad (10)$$

$$...$$

$$\forall t \in [\nu_k + \varphi_k, \infty), x(t) = f(u^k) \quad (11)$$

i.e. $x$ models $\tilde{x}$ on $[\nu_0 + \varphi_0, \nu_1] \vee [\nu_1 + \varphi_1, \nu_2] \vee ... \vee [\nu_k + \varphi_k, \infty)$.

b) For the unbounded delay model, we have a similar property satisfied with a), where we replace $\varphi_0, \varphi_1,...,\varphi_k$ with $\Theta_0, \Theta_1,...,\Theta_k$.

15.4 **Remark** A similar property with the one from 15.3 takes place if the input $u$ of $\Sigma \in [\tilde{\Sigma}]$ is given by 14.3 (4).

15.5 **Remark** The solution of the equations 15.2 (3), (4) is given by:

$$x_i(t) = x_i^0 \cdot \chi_{[0,\tau_i)}(t) \oplus c_i \cdot \chi_{[\tau_i,\infty)}(t), i = \overline{1,n} \quad (1)$$

In the fixed delay model, for example $x_i, i = \overline{1,n}$ are stable at $\tau_i, i = \overline{1,n}$ and this automaton may be considered, like in the Remark 14.6, to operate (by definition) in the fundamental mode.

## 16. The Unbounded Delay Model: Semi-Modularity. Synchronous-Like Autonomous Automata

16.1 **Informal definition**

a) [Beerel, Meng, 1991] Given the states $a,b$, the authors define the relations $R, F$ in the following manner. $a\, R\, b$: $b$ follows after $a$ with no interleaving states, $a\, F\, b$: $b$ follows after $a$ with any number of interleaving states. The transition $a \rightarrow b$ is defined to be semi-modular if "for any node $i$ which is enabled in state $a$ which does not change to its enabled value $a_i^{'}$ when the circuit goes to state $b$ is enabled in state $b$ and to the same value as in state $a$. A circuit is semi-modular with respect to state $\tau$ if every pair of states $a,b$ satisfying $\tau\, F\, a$ and $a\, R\, b$ has $b_i^{'} = a_i^{'}$ for each node $i$ satisfying $b_i = a_i \neq a_i^{'}$". $\tau$ is chosen to be the initial state.

b) [Lavagno, 1992] "A circuit is semi-modular if every excited signal becomes stable only by changing its value (i.e. not because one of the gate inputs has changed value)".

16.2 **Remark** The semi-modularity is the property of an "expected" transition to take place for any values $\tau_i > 0, i = \overline{1,n}$ - unbounded delay model, version 5.8 c) - and it is also the property of a system of having only such transitions. By "expected" transition it is understood that if the trajectory of the autonomous automaton $\Sigma$ takes at some time instant $t \geq 0$ the value $x(t)$, then we expect that some time instant $t' > t$ exists, depending on $\tau_1,...,\tau_n$ so that

$$x(t') = g(x(t)) \tag{1}$$

16.3 **Remark** The terminology of semi-modularity is inspired by the lattice theory. The author has included it because of the fact that it is usual in the asynchronous automata theory, even if the preferred terminology here is that of synchronous-like transition and synchronous-like system.

Semi-modularity is called by Grigore Moisil "the technical condition of good running". Many other authors call the semi-modular transitions of the asynchronous automata to be "hazard free".

16.4 **Definition** Let $g : \boldsymbol{B}_2^n \rightarrow \boldsymbol{B}_2^n, n \geq 1$ an arbitrary function. We define the *iterates* of $g$ to be the functions $g^k : \boldsymbol{B}_2^n \rightarrow \boldsymbol{B}_2^n, k \in \boldsymbol{N}$ given by the formulas:

$$g^0(x) = x \tag{1}$$

$$g^{k+1}(x) = g(g^k(x)), x \in \boldsymbol{B}_2^n \tag{2}$$

16.5 **Notations** We use from this moment the notations from Problem 10.10, where $\tau_i \leq M_i, i = \overline{1,n}$. $\boldsymbol{F}^0$ is the one from 10.12 and the solution $x$ is of the form 10.14 (1).

16.6 **Remark** The property

$$\forall k, g^{k+1}(x^0) \neq g^k(x^0) \tag{1}$$

under the condition of semi-modularity to be presented, is one of instability. Let us suppose the contrary, that

$$\exists k, g^{k+1}(x^0) = g^k(x^0) \tag{2}$$

is true. Then, the semi-modularity being satisfied, this last property is one of stability of the system; the sequence of vectors from $\boldsymbol{B}_2^n$: $x^0, g(x^0), g^2(x^0),...$ becomes constant starting with a certain rank and the constant (= the limit of the sequence) is an equilibrium point of $g$.

16.7 **Theorem** Let the automaton $\Sigma$ associated to the circuit $\tilde{\Sigma}$. The numbers $a_1^k,...,a_n^k \in \boldsymbol{B}_2$ are defined by:

$$g^k(x^0) \oplus g^{k+1}(x^0) = a_1^k \cdot \overset{-1}{\varepsilon} \oplus ... \oplus a_n^k \cdot \overset{-n}{\varepsilon}, k \in \boldsymbol{N} \tag{1}$$

i.e. the vectors $g^k(x^0), g^{k+1}(x^0)$ differ on the coordinates $i$ for which $a_i^k = 1$.

a) In the hypothesis of instability

$$\forall k \in \boldsymbol{N}, \exists i \in \{1,...,n\}, a_i^k = 1 \tag{2}$$

we define the numbers

$$\varphi_k = \max\{\tau_i \mid i \in \{1,...,n\}, a_i^k = 1\}, k \in \boldsymbol{N} \tag{3}$$

If

$$\forall k \in \boldsymbol{N}, \forall a_1,...,a_n \in \boldsymbol{B}_2, a_1 \le a_1^k,...,a_n \le a_n^k,$$

$$g^{k+1}(x^0) \ne g^k(x^0) \oplus a_1 \cdot \overset{-1}{\varepsilon} \oplus ... \oplus a_n \cdot \overset{-n}{\varepsilon} \Rightarrow$$

$$\Rightarrow g^{k+1}(x^0) = g(g^k(x^0) \oplus a_1 \cdot \overset{-1}{\varepsilon} \oplus ... \oplus a_n \cdot \overset{-n}{\varepsilon}) \tag{4}$$

then the natural indexes exist:

$$0 = q_0 < q_1 < q_2 < ... \tag{5}$$

and the numbers $\{t'_{q_k} \mid k \in \boldsymbol{N}\}$ from $\boldsymbol{F}^0$ defined by

$$t'_{q_0} = 0 \tag{6}$$

$$t'_{q_{k+1}} = \sum_{p=0}^{k} \varphi_p \tag{7}$$

so that the trajectory $x$ satisfies:

$$x(t'_{q_k}) = g^k(x^0), k \in \boldsymbol{N} \tag{8}$$

b) We suppose that the next condition of non-trivial stability is true:

$$\exists N \ge 1, \forall k \in \{0,...,N-1\}, \exists i \in \{1,...,n\}, a_i^k = 1 \tag{9}$$

$$a_1^N = ... = a_n^N = 0 \tag{10}$$

and we define

$$\varphi_k = \max\{\tau_i \mid i \in \{1,...,n\}, a_i^k = 1\}, k \in \{0,...,N-1\} \tag{11}$$

If

$$\forall k \in \{0,...,N-1\}, \forall a_1,...,a_n \in \boldsymbol{B}_2, a_1 \le a_1^k,...,a_n \le a_n^k,$$

$$g^{k+1}(x^0) \ne g^k(x^0) \oplus a_1 \cdot \overset{-1}{\varepsilon} \oplus ... \oplus a_n \cdot \overset{-n}{\varepsilon} \Rightarrow$$

$$\Rightarrow g^{k+1}(x^0) = g(g^k(x^0) \oplus a_1 \cdot \overset{-1}{\varepsilon} \oplus ... \oplus a_n \cdot \overset{-n}{\varepsilon}) \tag{12}$$

then the natural indexes exist:

$$0 = q_0 < ... < q_N \tag{13}$$

and the numbers $t'_{q_0},...,t'_{q_N}$ from $\boldsymbol{F}^0$ defined by

$$t'_{q_0} = 0 \tag{14}$$

$$t'_{q_{k+1}} = \sum_{p=0}^{k} \varphi_p \tag{15}$$

so that $x$ satisfies:

$$x(t'_{q_k}) = g^k(x^0), k \in \{0,...,N\} \tag{16}$$

$$\forall t \in [\Theta, \infty), x(t) = g^N(x^0) \tag{17}$$

i.e. $x$ models $\tilde{x}$ on $[\Theta, \infty)$, where we have noted

$$\Theta = \sum_{p=0}^{N-1} \max\{M_i \mid i \in \{1,...,n\}, a_i^p = 1\} \tag{18}$$

c) We suppose that the next condition of trivial stability is true:

$$a_1^0 = ... = a_n^0 = 0 \tag{19}$$

Then $x$ is given by 13.4 (2) and it models $\tilde{x}$ on $[0, \infty)$.

**Proof** a) We show that for all $k \in \boldsymbol{N}$, the functions

$$x_i(t) = g_i^k(x^0) \oplus a_i^k \cdot \chi_{[t'_{q_k} + \tau_i, t'_{q_{k+1}}]}(t), i = \overline{1,n} \tag{20}$$

satisfy EAAA when $t \in [t'_{q_k}, t'_{q_{k+1}}]$. Moreover, the next equalities are true:

$$\forall t \in [t'_{q_k}, t'_{q_{k+1}}), g(x(t)) = x(t'_{q_{k+1}}) = g^k(x^0) \oplus a_1^k \cdot \overline{\varepsilon}^1 \oplus ... \oplus a_n^k \cdot \overline{\varepsilon}^n = g^{k+1}(x^0) \tag{21}$$

16.8 **Definition** Let $t \geq 0$ and the numbers $a'_1,...,a'_n \in \boldsymbol{B}_2$ defined by:

$$x(t) \oplus g(x(t)) = a'_1 \cdot \overline{\varepsilon}^1 \oplus ... \oplus a'_n \cdot \overline{\varepsilon}^n \tag{1}$$

and we suppose that $x(t) \neq g(x(t))$. If for any binary numbers $a_1 \leq a'_1,...,a_n \leq a'_n$ so that $g(x(t)) \neq x(t) \oplus a_1 \cdot \overline{\varepsilon}^1 \oplus ... \oplus a_n \cdot \overline{\varepsilon}^n$ we have

$$g(x(t)) = g(x(t) \oplus a_1 \cdot \overline{\varepsilon}^1 \oplus ... \oplus a_n \cdot \overline{\varepsilon}^n) \tag{2}$$

then the transition $x(t) \to g(x(t))$ is called *synchronous-like*, or *semi-modular*.

By definition, the trivial transition $x(t) \to g(x(t)) = x(t)$ is synchronous-like.

An autonomous automaton where the transitions $g^k(x^0) \to g^{k+1}(x^0), k \in \boldsymbol{N}$ are synchronous-like is called synchronous-like (or semi-modular).

16.9 **Remark** The sense of Theorem 16.7 is of showing that a synchronous-like autonomous asynchronous automaton has for any $\tau_i > 0, i = \overline{1,n}$ a predictable trajectory.

If the vectorial sequence $g^k(x^0), k \in \boldsymbol{N}$ is divergent, then the time instants $\{t'_{q_k} \mid k \in \boldsymbol{N}\}$ from $\boldsymbol{F}^0$ exist, given by 16.7 (6), (7) for which 16.7 (8) is true and the automaton is unstable.

If the sequence $g^k(x^0), k\in \mathbf{N}$ is convergent (=constant from a certain rank), then the trajectory $x$ satisfies a similar property to the previous one expressed by the equations 16.7 (14), (15), (16) and the automaton is stable. Moreover, a set of the form $[\Theta,\infty)$ exists where $x$ equals $\lim\limits_{k\to\infty} g^k(x^0)$, being at the same time a model of $\tilde{x}$.

16.10 **Remark** The condition of synchronous-likeness is one of sufficiency. The transitions $g^k(x^0)\to g^{k+1}(x^0), k\in \mathbf{N}$ may result by the composition of other transitions, possibly elementary (this fact depends on the values of $\tau_i, i=\overline{1,n}$), but they always happen, for any values of $\tau_i>0, i=\overline{1,n}$, in exactly this order: $x$ runs through the values $g^0(x^0), g^1(x^0),...$ at some time instants $t'_{q0}, t'_{q1},...$

16.11 **Example** $x=(x_1,x_2,x_3)\in \mathbf{B}_2^3$,

$$x^0=(0,0,0) \qquad (1)$$

The generator function $g:\mathbf{B}_2^3\to \mathbf{B}_2^3$ and the trajectory of the system are the following:

| $x_1$ | $x_2$ | $x_3$ | $g_1$ | $g_2$ | $g_3$ |
|---|---|---|---|---|---|
| 0 | 0 | 0 | 0 | 1 | 1 |
| 0 | 0 | 1 | 0 | 1 | 1 |
| 0 | 1 | 0 | 0 | 1 | 1 |
| 0 | 1 | 1 | 1 | 1 | 1 |
| 1 | 0 | 0 | 0 | 0 | 0 |
| 1 | 0 | 1 | 1 | 0 | 0 |
| 1 | 1 | 0 | 1 | 0 | 0 |
| 1 | 1 | 1 | 1 | 0 | 0 |

table (2)

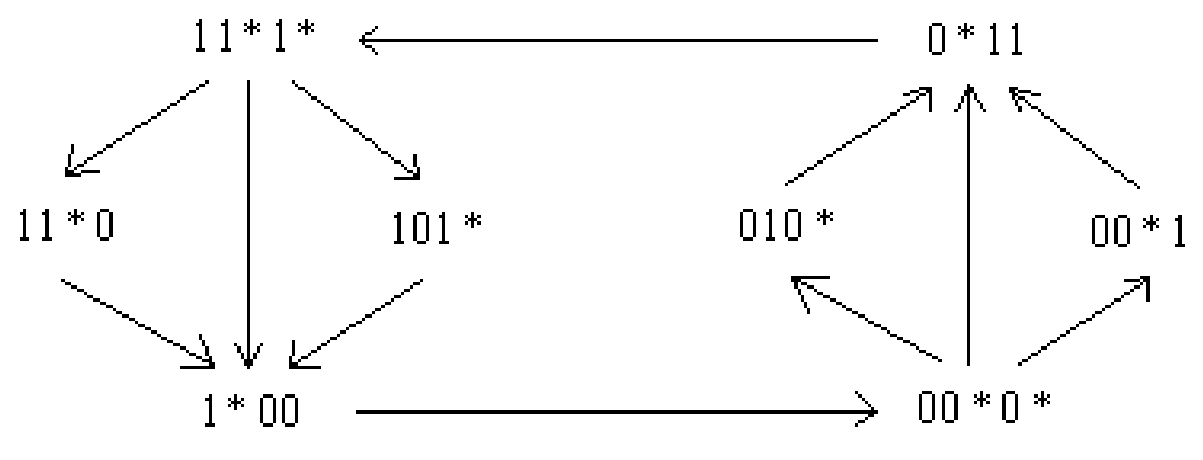

fig (3)

An asterisk notes, like before, the enabled coordinates. There are two races, so called *non-critical*: $000\to 011$ and $111\to 100$, because these transitions are independent on the race winner (they take place independently on the values of $\tau_1,\tau_2,\tau_3$).

16.12 **Remark** An autonomous synchronous-like asynchronous automaton enters a loop formed by the vectors $g^k(x^0), g^{k+1}(x^0),..., g^{k+p-1}(x^0)$ (because: $\mathbf{B}_2^n$ is a finite set and $g$ is a function):

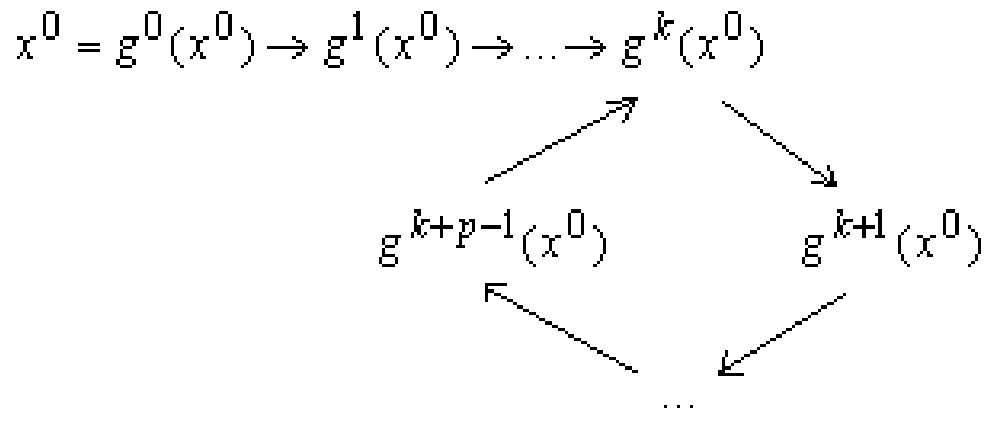

fig (1)

The hypothesis of instability demands that the loop contains more than one element, i.e. that $p \geq 2$. If the automaton is stable, then $p = 1$. If the stability is trivial, then $k=0$.

The situation $k=0$, $p \geq 2$ brings nothing new and the example 16.11 is of this type.

**17. Synchronous-Like Control Automata**

17.1 **Remark** The synchronous-like control automata combine the synchronous-like transitions with the fundamental mode of operation. These systems model the asynchronous circuits by generalizing from autonomous to non-autonomous the ideas from the previous paragraph.

17.2 **Definition** Let the arbitrary function $f : \boldsymbol{B}_2^n \times \boldsymbol{B}_2^m \to \boldsymbol{B}_2^n, n, m \geq 1$. The *iterates* of $f$ are defined to be the functions $f^k : \boldsymbol{B}_2^n \times \boldsymbol{B}_2^m \to \boldsymbol{B}_2^n, k \in \boldsymbol{N}$ given by:

$$f^0(x,u) = x \quad (1)$$

$$f^{k+1}(x,u) = f(f^k(x,u),u), x \in \boldsymbol{B}_2^n, u \in \boldsymbol{B}_2^m \quad (2)$$

17.3 **Notations** From this moment, the notations will be those from Problem 10.1 and we shall suppose in addition that $\tau_i \leq M_i, i = \overline{1,n}$ in the unbounded delay model 5.8 c''). It was noted with $\boldsymbol{F}$ the set from 10.5 and the trajectory $x$ of the non-autonomous automaton $\Sigma$ results to be of the form 10.7 (1).

17.4 **Theorem** $\Sigma$ is associated to $\tilde{\Sigma}$. The vectors $z^j \in B_2^n$ and the natural numbers $N_j \geq 1$, $j \in \boldsymbol{N}$ are defined so that the next conditions of non-trivial stability are fulfilled:

$$z^0 = x^0 \quad (1)$$

$$z^j = f^0(z^j, u^j) \neq f^1(z^j, u^j) \neq \ldots \neq f^{N_j}(z^j, u^j) = f^{N_j+1}(z^j, u^j) = \ldots \quad (2)$$

$$z^{j+1} = f^{N_j}(z^j, u^j) \quad (3)$$

whilst the family $a_{1j}^k, \ldots, a_{nj}^k \in \boldsymbol{B}_2$ is defined like this:

$$f^k(z^j, u^j) \oplus f^{k+1}(z^j, u^j) = a_{1j}^k \cdot \overline{\varepsilon}^1 \oplus \ldots \oplus a_{nj}^k \cdot \overline{\varepsilon}^n \quad (4)$$

$k \in \{0, \ldots, N_j - 1\}, j \in \boldsymbol{N}$. The real numbers $\varphi_k^j, \Theta_j > 0$ are defined in the next manner:

$$\varphi_k^j = \max\{\tau_i \mid i \in \{1, \ldots, n\}, a_{ij}^k = 1\} \quad (5)$$

$$\Theta_j = \sum_{p=0}^{N_j-1} \max\{M_i \mid i \in \{1,...,n\}, a_{ij}^p = 1\}, k \in \{0,...,N_j - 1\}, j \in \mathbf{N} \tag{6}$$

If

$$\forall k \in \{0,...,N_j - 1\}, \forall a_1,...,a_n \in \mathbf{B}_2, a_1 \le a_{1j}^k,...,a_n \le a_{nj}^k, \tag{7}$$

$$f^{k+1}(z^j,u^j) \ne f^k(z^j,u^j) \oplus a_1 \cdot \overline{\varepsilon}^{-1} \oplus ... \oplus a_n \cdot \overline{\varepsilon}^{-n} \Rightarrow$$

$$\Rightarrow f^{k+1}(z^j,u^j) = f(f^k(z^j,u^j) \oplus a_1 \cdot \overline{\varepsilon}^{-1} \oplus ... \oplus a_n \cdot \overline{\varepsilon}^{-n}, u^j)$$

and, moreover

$$\nu_{j+1} - \nu_j \ge \Theta_j \tag{8}$$

are true for all $j \in \mathbf{N}$, then for any $\tau_i > 0, i = \overline{1,n}$ the natural indexes exist

$$0 = q_0^0 < q_1^0 < ... < q_{N_0}^0 \le q_0^1 < q_1^1 < ... < q_{N_1}^1 \le q_0^2 < ... \tag{9}$$

and the numbers $\{t_{q_k^j} \mid k \in \{0,...,N_j\}, j \in \mathbf{N}\}$ from $\mathbf{F}$ defined by

$$t_{q_0^j} = \nu_j \tag{10}$$

$$t_{q_{k+1}^j} = \nu_j + \sum_{p=0}^{k} \varphi_p^j, k \in \{0,...,N_j - 1\}, j \in \mathbf{N} \tag{11}$$

so that the trajectory $x$ satisfies

$$x(t_{q_k^j}) = f^k(z^j,u^j) \tag{12}$$

$$\forall t \in [t_{q_{N_j}^j}, \nu_{j+1}], x(t) = z^{j+1} \tag{13}$$

for all $k \in \{0,...,N_j\}, j \in \mathbf{N}$. For any $\tau_i \in (0, M_i], i = \overline{1,n}$, $x$ models $\tilde{x}$ on the set $\bigvee_{j \in \mathbf{N}} [\nu_j + \Theta_j, \nu_{j+1}]$.

17.5 **Remark** Theorem 17.4, whose proof is obvious, states that: in the unbounded delay model (version 5.6 c")), if we have synchronous-like non-trivial transfers (17.4 (7)) and if the fundamental mode is true (17.4 (8)), then

a) $f^k(z^j,u^j) \to f^{k+1}(z^j,u^j)$ take place for any $\tau_i > 0, i = \overline{1,n}$, $k \in \{0,...,N_j - 1\}, j \in \mathbf{N}$

b) $x$ is equal to $z^{j+1}$ on the sets $[t_{q_{N_j}^j}, \nu_{j+1}], j \in \mathbf{N}$

c) $x$ models $\tilde{x}$ on $\bigvee_{j \in \mathbf{N}} [\nu_j + \Theta_j, \nu_{j+1}]$.

17.6 **Remark** Let us observe once again the manner in which the fundamental mode expresses a relation of compatibility between the environment (the speed of variation of the input) and the system's inertia + its ability to stabilize.

17.7 **Remark** An important special case of the discussed problems is the one when triviality appears under the form:

$$\exists j \in \boldsymbol{N}, u^j = u^{j+1} = u^{j+2} = ... \qquad (1)$$

Theorem 17.4 may now be rewritten in two versions:

a) $z^j = f^0(z^j, u^j) \neq f^1(z^j, u^j) \neq ... \neq f^{N_j}(z^j, u^j) = f^{N_j+1}(z^j, u^j) = ...$ (2)

b) $z^j = f^0(z^j, u^j) \neq f^1(z^j, u^j) \neq ... \neq f^k(z^j, u^j) \neq f^{k+1}(z^j, u^j) \neq ...$ (3)

the first version being the one of a stable automaton and the second, respectively the one of an unstable automaton.

17.8 **Remark** Theorem 17.4 gives the possibility of characterizing the asynchronous automaton $\Sigma$ by a synchronous (i.e. discrete time) automaton:

$$\boldsymbol{B}_2^n \times \boldsymbol{B}_2^m \ni (z^j, u^j) \mapsto z^{j+1} \in \boldsymbol{B}_2^n$$

In the original version from 17.4, the time set is $\boldsymbol{N}$ and in the version 17.7 a), the time set is finite.

17.9 **Example** of synchronous-like automaton. Let us take $n = 3$, $m = 1$ and

$$u(t) = \chi_{[0,\nu)}(t) \qquad (1)$$

By putting

$$x(t) = (x_1(t), x_2(t), x_3(t)), t \in \boldsymbol{R} \qquad (2)$$

we take in consideration the automaton

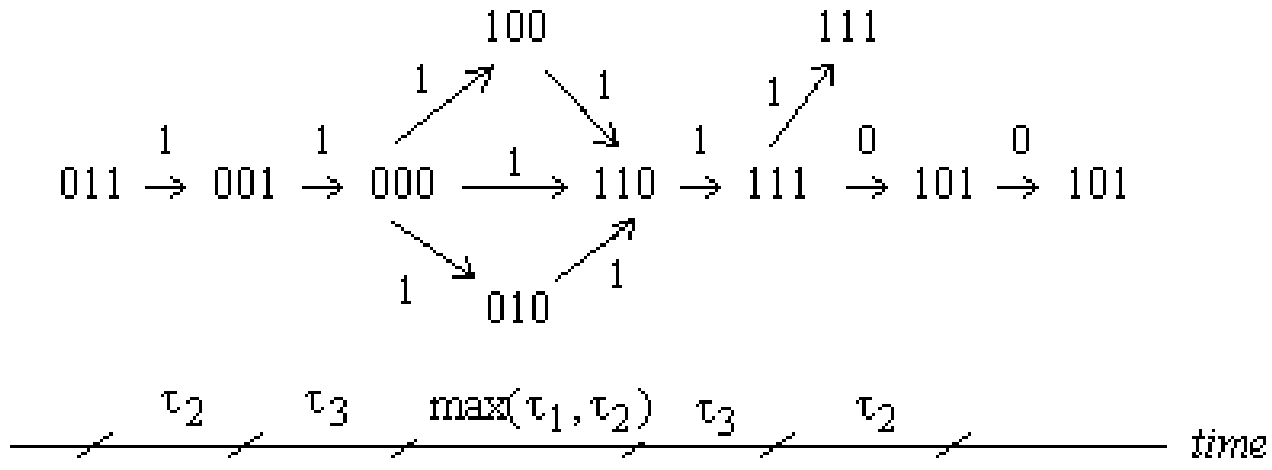

fig (3)

The numbers 0,1 put above the arrows show the value of the input that produces the synchronous-like transfer. The conditions 17.4 (8) are given by the inequality:

$$\nu \geq M_2 + 2M_3 + max\{M_1, M_2\} \qquad (4)$$

## 18. Classical and Linear Time Temporal Logic of the Propositions: Semantics

18.1 **Definition** The binary variables (or Boolean variables) $x_1,...,x_n \in \boldsymbol{B}_2$ will also be called the *atomic propositions* of the classical logic of the propositions CLP.

18.2 **Definition** The Boolean functions $h : \boldsymbol{B}_2^n \to \boldsymbol{B}_2$ are also called *formulas* of CLP.

18.3 **Remark** Defining the formulas as Boolean functions identifies the logically equivalent formulas and this is convenient in semantics.

18.4 **Remark** We observe, by following [Reghis, 1981] that no formal text, as long as it can be, needs an infinite number of signs and consequently it can be written even if the list of the binary variables $x_1,...,x_n$ is finite and sufficiently large. This creates no loss of generality.

18.5 **Remark**, the informal definition of the CLP semantics. The semantic approach of CLP answers the question: in the interpretation $I$ that gives the variable $x=(x_1,...,x_n)$ the constant value $x^0=(x_1^0,...,x_n^0)\in \boldsymbol{B}_2^n$ do we have $h(x^0)=1$?

18.6 **Definition** If $h(x^0)=1$, we say that $h$ *is satisfied in* $I$ or that it *holds in* $x^0$ and we note this fact with $x^0 \models h$.

18.7 **Definition** If $h=1$ (the constant function), we say that it is a *tautology* and we note this fact by $\models h$.

18.8 **Remark** Now we shall pass to the temporal computation of the Boolean functions, i.e. to the temporal logic, in two variants: a) continuous time and b) discrete time. The coexistence of the continuous time with the discrete time has been underlined many times in this work.

18.9 **Remark** We shall suppose in this paragraph the validity of the fixed delay model, thus we associate the fixed delay model to the linear time temporal logic.

18.10 **Definition** a) The functions $x_1,...,x_n \in Diff_+$ are also called the *atomic propositions* of the linear time temporal logic of the propositions, LTL

b) Similarly, the functions $x_1,...,x_n : \boldsymbol{F} \to \boldsymbol{B}_2$ are also called the *atomic propositions* of LTL, where $\boldsymbol{F}=\{t_k \mid k\in \boldsymbol{N}\}$ is the countable time set.

18.11 **Remark** We shall interpret $x_1,...,x_n$ to be the coordinates of the state of an automaton $\Sigma$ - even if, in general, they are arbitrary - and then $\boldsymbol{F}$ is the alternative time set that has been defined at 10.5. The relation from a systemic point of view between the real time and the discrete time is the one from 10.7.

18.12 **Definition** The set of the formulas of LTL is defined in the next manner.

a) In the continuous version, the formulas are functions $Diff_+^{(n)} \to Diff_+$.

a.1) the Boolean functions $h:\boldsymbol{B}_2^n \to \boldsymbol{B}_2$ induce the formulas

$$h(x)(t)=h(x(t))\cdot \chi_{[0,\infty)}(t) \tag{1}$$

These formulas have the same names like $h$: reunion, intersection etc.

a.2) if $h$ is a formula, then $h^-$ is also a formula, called the *left limit of* $h$, which is defined by:

$$h^-(x)(t)=h(x)(t-0) \tag{2}$$

a.3) if $g,h$ are formulas, then $hUg$ is a formula, called $h$ *until* $g$. The definition is:

$$(hUg)(x)(t)=\bigcup_{t'\geq t} g(x)(t')\cdot \bigcap_{\xi\in[t,t')} h(x)(\xi) \tag{3}$$

a.4) all the LTL formulas are defined by a.1),…,a.3), where $x\in Diff_+^{(n)}$ and $t\in \boldsymbol{R}$.

b) in the discrete version, the formulas are functions that associate to a sequence of vectors $x(t_k)=(x_1(t_k),...,x_n(t_k))$, respectively a binary sequence $h(x)(t_k)$, where $t_k\in \boldsymbol{F}, k\in \boldsymbol{N}$.

b.1), b.3), b.4) are similar to a.1), a.3), a.4)

b.2) if $h$ is a formula, then $Xh$ is a formula (noted sometimes with $Oh$), called *next* $h$. The definition is:

$$Xh(x)(t_k) = h(x)(t_{k+1}) \qquad (4)$$

18.13 **Example** The modulo 2 sum and the intersection bring functions from $Diff_+$ to functions from $Diff_+$. The complement brings the null function to the function $\chi_{[0,\infty)}$.

18.14 **Remark** The informal definition of the LTL semantics is made similarly to 18.5.

18.15 **Remark** It is interesting that semantic approach of LTL answering the question: in the interpretation that gives the argument $x$ of $h$ the constant value, noted with the same symbol, representing the solution of EAA and fixes the time to $t \geq 0$, do we have that $h(x)(t) = 1$?

18.16 **Definition** If the answer to the previous question is positive, we say that $h$ *is satisfied at* $t$ ($x$ is kept in mind), or that it *holds at* $t$ and we note this fact with $t \models h$. In the discrete version, the equation $h(x)(t_k) = 1$ is noted with $k \models h$. In both versions, instead of $0 \models h$ we write $\models h$.

18.17 **Remark** The semantics that we use here is called *floating* and it differs from the *anchored* semantics by the fact that the latter refers only to statements of the type $\models h$.

18.18 **Remark** In [Alfaro, Manna, 1995] it is mentioned the fact that in their theory the operator $X$ is missing. The continuous semantics of this operator is rather given by the equation $h(x)(t-0) = 1$, noted $t \models h^-$, then by $h(x)(t+0) = 1$, noted $t \models h^+$, as it would seem to be normal; the realizable functions are right continuous (see 2.23 (2), for example) and the operator $h^+ = h$ is of null effect in the non-anticipative reasoning. Dually, for the anticipative asynchronous automata together with the appropriate reasoning, the operator $h^- = h$ is of null effect.

It is likely that in the cited paper that relates the discrete time to the continuous time $X$ is missing because reasoning there did not start from behind systems theory; when trying to make the continuous analogue of $X$ be the "right limit" operator, a failure results.

18.19 **Remark** $U$ gives the possibility of defining the unary connectors *Always*, *Henceforth*, or *Necessity* $G$, respectively *Sometimes*, *Eventually*, or *Possibility* $F$. For example if in 18.12 (3) $h = 1$ (to be understood that $h$ is equal to the unit of $Diff_+$, which is $\chi_{[0,\infty)}$) then:

$$(Fg)(x)(t) = (1Ug)(x)(t) = \bigcup_{t' \geq t} g(x)(t') \qquad (1)$$

and $\bigcup_{t' \geq t} g(x)(t') = 1$ is noted with $t \models Fg$, when $x$ is the solution of EAA. We read: "it is possible starting with $t$ that $\Sigma$ has the property $g$". This is the connection with the modal logic.

18.20 **Remark** Alfaro and Manna mention in the syntax of their temporal logic the *age function* $\Gamma$: "*for a formula* $h$*, at any point in time, the term* $\Gamma(h)$ *denotes for how long in the past* $h$ *has been continuously true*". Let us remark that such an idea occurs in EAA in the term $\overline{\bigcup_{\xi \in (t-\tau_i, t)} D^- f_i(x(\xi), u(\xi))}$: if $f_i(x(\xi), u(\xi))$ is *constant* on the interval $(t - \tau_i, t)$, its derivative is null on this interval, the reunion is null also and its complement is unitary; this is a necessary condition (of inertiality) to have $D^{-}x_i(t)=1$. Thus the asynchronous automata make use of $n$ age functions $\Gamma_i$ which replace in their definition 'continuously true' with 'continuously constant' and limit coordinatewise the memory of the automaton to $\tau_i$ time units.

## 19. Branching Time Temporal Logic of the Propositions: Semantics

19.1 **Remark** The solution *x* of EAA, expressed by Theorem 10.9 depends on $\tau_1,...,\tau_n$ in both temporal variants, real time and discrete time, in the sense that $\{t_k \mid k\in \boldsymbol{N}\}$ and the sequence $\{x^k \mid k\in \boldsymbol{N}\}$ both depend on these parameters, that are subject to the restrictions 5.8.

19.2 **Notation** We note with $\pi:\boldsymbol{B}_2^{m+n}\rightarrow \boldsymbol{B}_2^n, m\geq 0, n\geq 1$ the projection on the last $n$ coordinates.

19.3 **Proposition** $\{x^k \mid k\in \boldsymbol{N}\}\subset \pi(dom\,\Gamma)$, where $\Gamma$ (see 12.5) is the set of the transitions.

19.4 **Definition** The delay model of the delay element and the delay itself are called *pure chaos,* if the SINLF family $\{\psi_j \mid j\in \boldsymbol{N}\}$ and the positive numbers $\tau(\psi_j), j\in \boldsymbol{N}$ exist (see 5.10) with

$$\tau(t)=\sum_{j\in \boldsymbol{N}}\tau(\psi_j)\cdot\chi_{[\psi_j,\psi_{j+1})}(t), t\geq 0 \qquad (1)$$

19.5 **Remark** We note with $\boldsymbol{F}=\{t_k \mid k\in \boldsymbol{N}\}$ the SINLF family that is obtained by counting the elements of the set

$$\{\nu_k+p_1\cdot\tau_1(\psi_j)+...+p_n\cdot\tau_n(\psi_j) \mid [\psi_j,\psi_{j+1})\wedge[\nu_k,\nu_{k+1})\neq\emptyset, k, j, p_1,...,p_n\in \boldsymbol{N}\}$$

in a strictly increasing order, the way we did in the situations from 5.8, and we note with $x$, see 10.7 (1), the solution of EAA'. By definition, EAA' coincides with EAA, except for the fact that $\tau_1,...,\tau_n$ are not constant, but piecewise constant, i.e. pure chaos.

The study of EAA' was not made, but we accept the fact that these equations have a unique solution, that is expressed by a theorem similar to 10.9 and that 19.3 is true in this situation too.

19.6 **Definition** When $\tau_1,...,\tau_n$ vary in one of the manners from 5.8, 19.4, the appropriate solution x is called a *path.* In the continuous version of the time, a path is a function $x\in Real^{(n)}$ and in the discrete version of the time, a path is a sequence $(x(t_k))_{k\in \boldsymbol{N}}$.

19.7 **Remark** Of course that the "branching time" resulting from the existence of several paths does not mean the presence of $x$ on several branches of the time set, but choosing one of them.

19.8 **Notation** We note with $Path$ the set of the paths of $\Sigma$ (the same notation for two sets, one for the continuous time and one for the discrete time).

19.9 **Remark** The branching time temporal logics of the propositions, for example $BT, BT^+, UB, UB^+, CTL, CTL^+$, $CTL^*$ [Gupta, 1991], $\forall CTL, \exists CTL, \forall CTL^*, \exists CTL^*$ [Vardi, 1994] have common features. In their syntax appear like at the linear time temporal logic the Boolean connectors, as well as the temporal connectors $F, G, X, U$ regarded to be *time quantifiers.* In addition we have the *path quantifiers* $A$ and $E$.

19.10 **Definition** The set of the formulas of the branching time temporal logic of the propositions is defined like this.

a) In the continuous version, the formulas are functions $Diff_+^{(n)} \to Diff_+$

a.1) if $h$ is an LTL formula, then it is a formula of the branching time temporal logic

a.2) if $h$ is a formula of the branching time temporal logic, then $Ah$ and $Eh$ are such formulas, called *for all paths* $h$, respectively *for some path* $h$. The definitions are:

$$Ah(x)(t) = \bigcap_{x \in Path} h(x)(t) \quad (1)$$

$$Eh(x)(t) = \bigcup_{x \in Path} h(x)(t) \quad (2)$$

a.3) All the formulas of the branching time temporal logic are given by a.1), a.2).

b) In the discrete version, the situation is similar.

19.11 **Remark** The functions $Ah, Eh$ depend on $t$, but they do not depend on $x$. From here we get the manner in which the semantical notations $t \models Ah, k \models Ah$ etc, written similarly to 18.16, must be interpreted, see also 18.15.

## 20. Conclusions

The study of the asynchronous circuits is based on intuition and on a bibliography which is often insufficiently formalized. The paper gives a model for these circuits. Some of the topics that we have dealt with are: transitions, stability, the fundamental mode of operation, the special case of the combinational automata, semi-modularity and synchronous-like automata, connections with temporal logic, whose formulas present, in continuous or discrete time, the properties of the system.